\documentclass[a4paper,11pt]{scrartcl}
\pdfoutput=1

\makeatletter
\DeclareOldFontCommand{\rm}{\normalfont\rmfamily}{\mathrm}
\DeclareOldFontCommand{\sf}{\normalfont\sffamily}{\mathsf}
\DeclareOldFontCommand{\tt}{\normalfont\ttfamily}{\mathtt}
\DeclareOldFontCommand{\bf}{\normalfont\bfseries}{\mathbf}
\DeclareOldFontCommand{\it}{\normalfont\itshape}{\mathit}
\DeclareOldFontCommand{\sl}{\normalfont\slshape}{\@nomath\sl}
\DeclareOldFontCommand{\sc}{\normalfont\scshape}{\@nomath\sc}
\makeatother

\usepackage{hyperref}
\usepackage{xspace}
\usepackage{pslatex}
\usepackage{amsmath}
\usepackage{txfonts}
\usepackage{graphicx}
\usepackage{cancel}
\usepackage{amssymb}
\usepackage{color}
\usepackage{authblk}
\usepackage{array,multirow}

\graphicspath{{figs/}}

\def\mt{m_t}

\def\nn{\nonumber}

\def\min{{\textrm min}}

\def\NLO{{\textrm{NLO}}}
\def\Born{{\textrm{Born}}}
\def\Refs#1{refs.~\cite{#1}}
\def\Ref#1{ref.~\cite{#1}}
\def\Eq#1{eq.~(\ref{#1})}
\def\Fig#1{fig.~\ref{#1}}
\def\FIG#1{Fig.~\ref{#1}}

\def\Kallen#1{\lambda\left(#1\right)}

\def\smin{\ensuremath{s_{\text{min}}}\xspace}
\def\GeV{\ensuremath{\text{GeV}}\xspace}
\def\muR{\ensuremath{\mu_R}\xspace}
\def\muF{\ensuremath{\mu_F}\xspace}
\def\MEM{Matrix Element Method\xspace}
\def\PSS{Phase Space Slicing Method\xspace}
\def\PSSP{Phase Space Slicing parameter\xspace}

\def\had{{\mbox{\scriptsize had}}}
\def\incl{{\mbox{\scriptsize incl.}}}
\def\excl{{\mbox{\scriptsize excl.}}}

 \numberwithin{equation}{section}
 \numberwithin{figure}{section}
\titlehead{\hfill HU-EP-17/28, arXiv:1712.04527 }
 \title{The \MEM at next-to-leading order QCD for hadronic collisions: Single top-quark production at the LHC as an example application}
 \author[]{Till Martini\thanks{Till.Martini@physik.hu-berlin.de} }
 \author[]{Peter Uwer\thanks{Peter.Uwer@physik.hu-berlin.de}}
 \affil[]{Humboldt-Universit{\"a}t zu Berlin, Institut f{\"u}r Physik,
   Newtonstra{\ss}e 15, 12489 Berlin, Germany}

\begin{document}
\maketitle
\begin{abstract}
  Recently, a general algorithm to
  extend the \MEM (MEM) by taking into account next-to-leading-order (NLO) corrections in
  quantum chromodynamics (QCD) has been presented. In this article, the algorithm is
  applied to the most general case that coloured partons are encountered in the initial as well as the final state. This represents a substantial extension compared to previous work. As a concrete example, the production of single top quarks at the LHC is studied. We present in detail the generation of unweighted events following the NLO predictions. By treating these events as the result of a toy experiment, we show the first proof-of-principle application of the \MEM at NLO QCD for hadronic jet production. As an illustration, we study the determination of the top-quark mass. We find
  that---apart from elevating the powerful MEM to a sound theoretical foundation at NLO---the inclusion of the NLO corrections can lead to sizeable
  effects compared to the \MEM relying on leading-order predictions only. Furthermore, we find that the incorporation of the NLO corrections is mandatory to
  obtain reliable estimates of the theoretical uncertainties.  In
  addition, this work shows that measuring the top-quark mass using the MEM in
  single top-quark production offers an interesting alternative to mass
  measurements in top-quark pair production.
\end{abstract}
\vfill
\newpage
\tableofcontents

\section{Introduction}
In recent years multivariate methods have been proven to be
instrumental for the data analysis at the Large Hadron Collider (LHC).
Neuronal networks (NN), boosted decision trees (BDT) and the \MEM
(MEM) as specific examples belong to the standard techniques
employed in the LHC analysis today. Among these methods the \MEM is of
particular interest since the approach provides a very clean
statistical interpretation based on a direct link between theory and event reconstruction. It can be used for signal-background
discrimination as well as for parameter extraction. The \MEM as
introduced in \Ref{Kondo:1988yd,Kondo:1991dw} relies on the assumption
that the differential cross section can be interpreted as the probability
distribution to observe a particular event signature. Under this
assumption, the cross section is used to define a likelihood
function. Parameter determinations can be performed by maximising
this likelihood with respect to the model parameters used in the
theoretical predictions.  Signal-background discrimination can be
performed by defining likelihood ratios for signal+background and the
background only hypotheses. Both aspects have been pioneered in
the top-quark analysis at the Tevatron. In the early days of the top-quark
discovery the \MEM was used to determine the top-quark mass based on
$\mathcal{O}(100)$ events only
\cite{Abbott:1998dn,Abazov:2004cs,Abulencia:2006mi}. See \Ref{Fiedler:2010sg} for an introduction to the \MEM in the context of top-quark mass measurements. In the
single top-quark studies the method was instrumental to separate the
signal from the overwhelming background (see e.g. \Ref{Giammanco:2017xyn}). More recent
applications can be found in
\Refs{FerreiradeLima:2017iwx,Gritsan:2016hjl,Englert:2015dlp,Aad:2015upn,Khachatryan:2015ila,Artoisenet:2013vfa,Artoisenet:2013puc,Anderson:2013afp,Andersen:2012kn}.
An automation of the \MEM at leading order is presented in \Ref{Artoisenet:2010cn}.

The strength of the \MEM, namely offering a clean statistical
interpretation, also points to a potential weakness: The method will
fail to produce reliable results if the theoretically calculated cross
section does not provide a decent description of the underlying
probability distribution. In this case the statistical analysis shows
that in general parameter extraction with the \MEM no longer provides an unbiased estimator. In principle, this does
not preclude the usage of the \MEM for parameter extraction, since the
bias can be removed by an additional calibration procedure. However,
it would make the application of the \MEM less attractive because the
calibration introduces additional uncertainties and might rely on auxiliary modelling. Facing the
increasing precision current experiments are aiming for, it is thus
mandatory to include higher-order QCD corrections to the cross section predictions
used in the \MEM. However, so far the \MEM as applied in the experimental
analysis relies on leading-order cross section predictions only. In recent years
various attempts have been made to extend the \MEM to NLO.  In
\Ref{Alwall:2010cq} the effect of QCD radiation is studied as a first
step in this direction. In \Refs{Soper:2011cr,Soper:2014rya} the information from
the hard LO matrix element is combined with a parton shower. In
\Ref{Campbell:2012cz} a method is presented to consider NLO QCD corrections within the \MEM for the production of uncoloured final states. A first application is studied in
\Ref{Campbell:2013hz}.  In \Ref{Campbell:2013uha} the inclusion of
hadronic production of jets by mapping NLO and LO jets with a
longitudinal boost along the beam axis to remove the unbalanced
transverse momentum is investigated. However, no general algorithm to
include NLO QCD corrections for the hadronic production of coloured
partons within the \MEM is presented.  Only very recently a complete
algorithm to systematically include NLO QCD corrections within the
\MEM has been presented in \Ref{Martini:2015fsa}. As a proof of
concept the Drell-Yan process and top-quark pair production in $e^+e^-$-
annihilation has been studied in \Ref{Martini:2015fsa}. While
\Ref{Martini:2015fsa} relies on a well-motivated modification of the
clustering prescription used within the jet algorithm, it was shown
in \Ref{Baumeister:2016maz} how the method can be extended avoiding
the modification of the clustering prescription. However, the
extension presented in \Ref{Baumeister:2016maz} relies on the
numerical solution of a system of non-linear equations and the proof
of the feasibility of this approach in practical applications is still
missing.

The examples studied in \Ref{Martini:2015fsa} were chosen to offer
sufficient complexity to test all the aspects of the proposed
algorithm while still being simple enough to not pose any challenge to the numerical implementation. However, phenomenologically wise the two examples are
of limited interest although the studies of top-quark pair production
in $e^+e^-$-annihilation may provide useful information for a future
linear collider. In this article we apply the \MEM
at NLO QCD to hadronic single top-quark production as a particular example for the general case that coloured partons occur in the initial as well as the final state.  Because of the large
backgrounds single top-quark production is very challenging and the
usage of the \MEM is well motivated. Although top-quark production
in hadronic collisions is dominated by top-quark pair production,
single top-quark production plays a central role in top-quark physics
at the LHC, since it provides complementary information. In
particular, single top-quark production is a unique source of polarised
top-quarks and allows detailed tests of the $V-A$ structure of the
coupling to the $W$-boson and gives access to the
Cabbibo-Kobayashi-Maskawa matrix element $V_{tb}$.

In leading order three different
production modes are distinguished in the Standard Model (SM):
\begin{eqnarray}
  \label{eq:topcht}
  qb\rightarrow tq',& q^2_W<0  &\textrm{$t$-channel production},\\
  \label{eq:topchWt}
  gb\rightarrow tW^-,&q^2_W=m^2_W &\textrm{$Wt$-channel},\\
  \label{eq:topchs}
  qq'\rightarrow t\bar{b},& q^2_W>m^2_t &\textrm{$s$-channel production}.
\end{eqnarray}
At the LHC single top-quark production is dominated by the
$t$-channel followed by the $Wt$-channel. The $t$-channel 
total cross section has been precisely measured at the LHC at
7, 8 and 13 TeV by both CMS and ATLAS
\cite{Chatrchyan:2012ep,Aad:2014fwa,
  Khachatryan:2014iya,ATLAS:2014dja,Sirunyan:2016cdg,Aaboud:2016ymp,
  CMS:2015jca,CMS:2014ika,CMS:2016xnv}.

The $s$-channel production gives only a very small contribution at the
LHC. Because of the complicated backgrounds the signal is difficult to
extract and only recently evidence has been reported by the ATLAS
collaboration \cite{Aad:2015upn}. In \Ref{Aad:2015upn} the usage of the \MEM has been proven
instrumental to extract the signal. The independent measurement of
$t$- and $s$-channel production provides useful information to
constrain physics beyond the Standard Model (see for example \Ref{Giammanco:2017xyn}
for an overview of recent results).  Anticipating further experimental
progress, we present in this article the application of the \MEM to
$s$- and $t$-channel production including NLO QCD corrections, relying
on existing results for the NLO corrections.  Next-to-leading order results for the $s$- and $t$-channel have been known for quite a while
(see e.g. ref. \cite{Harris:2002md,Cao:2004ky} and are available in
numerical implementations like MCFM (see ref.  \cite{Campbell:2010ff})
or HATHOR (see ref. \cite{Kant:2014oha}).

In this article we apply the algorithm presented in ref. \cite{Martini:2015fsa}
to single top-quark production at the LHC. We present the
first application of the MEM at NLO to the hadronic production of jets
originating from coloured final state partons. As a proof of concept,
we generate unweighted $s$- and $t$-channel single top-quark events 
following the NLO predictions (with
and without a veto on an additional resolved jet) and apply the MEM at
NLO to reproduce the top-quark input mass from these events. 

The
article is organised as follows. In section~\ref{sec:IntroMEM} we give
a brief review of the \MEM. In particular, we focus on aspects
relevant for incorporating NLO QCD corrections. Furthermore, we give
a short description of the approach presented in \Ref{Martini:2015fsa} to extent
the \MEM to NLO accuracy. In section~\ref{sec:unweighted-events} we
discuss the generation of unweighted events following the NLO cross
section predictions. By comparing with a traditional parton level Monte
Carlo we also validate the entire procedure. We study an inclusive as
well as an exclusive event definition. The exclusive event definition
in which we veto additional jet emission is used to allow a more
detailed test of the procedure. Interpreting the unweighted events
generated in section~\ref{sec:unweighted-events} as the result of a
pseudo-experiment, we illustrate the application of the \MEM including
NLO QCD corrections in section~\ref{sec:MEMappl}. As a detailed test
of the method, we simulate a measurement of the top-quark mass. The
main results are summarised in section~\ref{sec:concl}.

\section{The \MEM}
\label{sec:IntroMEM}

\subsection{The \MEM in the Born approximation}
\label{sec:MEM-Basics}
To set up the notation used in this article, we briefly review the
essence of the \MEM in the Born approximation. A detailed description
focusing on top-quark mass measurements with the \MEM is given for
example in \Ref{Fiedler:2010sg}. The basic idea is to treat the
normalised differential cross section calculated from the squared
matrix elements as the probability distribution to observe a specific
event in the final state. We may think of a parton event as a
collection of partonic momenta. However, for the following discussion
it turns out to be useful to generalise the event definition to an
arbitrary collection of variables used to describe the event. This set
of variables which we may collect into a vector $\vec{x}$ does not
need to be complete in the sense that all available information about
the event can be reconstructed. The dimension $r$ of the vector
$\vec{x}$ may thus be smaller than the number of independent variables
$3n-4$, where $n$ denotes the number of outgoing partons.  In the
spirit of the \MEM the probability distribution for the production of
an event described by $\vec{x}$ is then described by the
(parameter-dependent) differential cross section
\begin{equation}
  P(\vec{x},\vec{\alpha}) = 
  {1\over \sigma(\vec{\alpha})} {d\sigma(\vec{\alpha}) \over dx_1\ldots dx_r}.
\end{equation}
The vector $\vec{\alpha}$ collects all the parameters e.g.  couplings
and masses on which the cross section depends.  The normalisation
factor $\sigma(\vec{\alpha})$ is given by the (parameter-dependent)
fiducial cross section:
\begin{equation}
  \sigma(\vec{\alpha}) = \int dx_1\ldots dx_r\; 
  {d\sigma(\vec{\alpha}) \over dx_1\ldots dx_r}.
\end{equation}
The probability distribution $P(\vec{x},\vec{\alpha})$ is thus
normalised to one.  The differential cross section is defined
schematically through
\begin{eqnarray}
  \label{eq:partonic-diff-xs}
  {d\sigma(\vec{\alpha})\over dx_1\ldots dx_r} &=&   
  \sum_{i,j} \int d\hat x_1 d\hat x_2 F_{i/{H_1}}(\hat x_1)
  F_{j/{H_2}}(\hat x_2)\nn \\
  &\times& {1\over 2s_\had \hat x_1 \hat x_2} \int dR_n(\vec{y},\vec{\alpha})) 
  |{\cal M}_{ij}(\vec{y},\vec{\alpha})|^2 
  \delta(\vec{x}-\vec{x}(\vec{y},\hat x_1, \hat x_2)).
\end{eqnarray}
As usual, the $ F_{i/{H_1}}(x)$ denote the parton distribution
functions and $s_\had$ the hadronic centre-of-mass energy squared. The
sum runs over all possible parton configurations $(ij)$ in the initial
state. If more than one sub-process contributes to a given event
signature (e.g. background) the right hand side of
\Eq{eq:partonic-diff-xs} needs to be extended to include a sum over
all contributing sub-processes.  The partonic variables used to
parameterise the $n$-parton phase space measure $dR_n$ as well as the
squared matrix elements $|{\cal M}_{ij}|^2$ are collected in
$\vec{y}$.  The function $\vec{x}(\vec{y},\hat x_1,\hat x_2)$ maps the
integration variables to the variables $\vec{x}$ used to describe the
partonic event. In experiments we do not observe partons but hadrons
and leptons. Using again a collection of variables $\vec{{\cal X}}$ to
describe the hadronic event as observed in the experiment the
probability distribution reads:
\begin{equation}\label{eq:probdist}
   P(\vec{\cal X},\vec{\alpha}) = {1\over \sigma(\vec{\alpha})} 
   \int dx_1\ldots dx_r {d\sigma(\vec{\alpha})\over dx_1\ldots dx_r}
   W(\vec{x},\vec{\cal X}),
\end{equation}
where the `transfer function' $W(\vec{x},\vec{\cal X})$ describes
the probability to observe a partonic event $\vec{x}$ as an hadronic
event $\vec{\cal X}$. The transfer functions have to be normalised
\begin{equation}
\int d\vec{\cal X}\; W(\vec{x},\vec{\cal X}) =1
\end{equation}
in order to ensure the correct normalisation of the probability
density.  In most applications one assumes that $W$ factorises into a
product of transfer functions describing the transition of individual
partons and leptons. For leptons where the energy and the direction of
flight can be measured with good accuracy the transfer functions are
often modelled using narrow Gaussian distributions. For unobserved
particles the transfer function is replaced by $1$ leading to an
integration of the related variables over the full phase space and the
aforementioned reduction of variables used to describe the
differential cross section.  For simplicity, we assume in the
following an ideal detector where the transfer functions are replaced
by $\delta$-functions. Note that the transfer functions model a
variety of different effects. They form an interface between
theoretical predictions and experimental observations and thus bridge
the gap between the two. Their impact may be reduced by improving the
theoretical modelling (e.g. considering $W^+W^-b\bar{b}$ instead of
$t\bar{t}$ final states) or including higher-order corrections in the
predictions. Furthermore, it is conceivable that the experimental data
could be unfolded to some intermediate level further reducing the
impact of the transfer functions. Using $\delta$-functions may thus
provide a useful first approximation.  As far as the conceptual
aspects of the method used in this article are concerned, the
inclusion of non-trivial transfer functions will not change any of the
arguments presented here. Non-trivial transfer function would only add
additional integrations. In practice however, the inclusion of more realistic
transfer functions may lead to additional technical problems as far as the
numerical integration is concerned. This is well known from the
application of the \MEM in leading-order. We thus consider the results
presented in this article only as a first step towards a more
realistic analysis. Additional technical
improvements may be required to include transfer functions
as determined in the experimental analysis.

We now briefly review the possibilities to use the \MEM for
signal-background discrimination and parameter estimation. In
addition, we introduce the concept of the extended likelihood and
collect some useful properties of estimators extracted with the \MEM.
\paragraph{Signal-background discrimination}
As mentioned before,
\Eq{eq:partonic-diff-xs} may include signal as well as background
processes. In \Ref{Cranmer:2006zs} it has been argued that calculating the
probability distribution independently for signal and signal plus
background an optimal discriminator can be constructed using the
ratio.  Thus, the \MEM makes optimal use of the
information contained in an event sample. Recent experimental work on
establishing $s$-channel single top-quark production provides a
nice example \cite{Aad:2015upn} of the usefulness of the \MEM in establishing ``signals''. 
\paragraph{Parameter extraction}
However, this is not the only
application of the \MEM. Since $P(\vec{\cal X},\vec{\alpha})$ in \Eq{eq:probdist} depends
on the theory parameters $\vec{\alpha}$ the \MEM can also be used to
determine the model parameters. For a given event sample $\{\vec{\cal
  X}_1,\ldots, \vec{\cal X}_N\}$ the likelihood
\begin{equation}
  \label{eq:likelihood}
  {\cal L}(\vec{\alpha}) = \prod_{i=1}^N  P(\vec{\cal X}_i,\vec{\alpha})= \prod_{i=1}^N\int d\vec{x}\;{1\over \sigma(\vec{\alpha})}\; {d\sigma(\vec{\alpha})\over d\vec{x}}\;
   W(\vec{x},\vec{\cal X}_i)
\end{equation}
is a function of the model parameters. Maximising the likelihood with respect to the model parameters $\vec{\alpha}$
yields an estimator for the true parameters. This
technique has been applied with great success to the top-quark mass
determination at the Tevatron collider (see \Refs{Abbott:1998dn,Abazov:2004cs,Abulencia:2006mi}).  
\paragraph{Extended likelihood}
The likelihood as defined in \Eq{eq:likelihood} only depends on the normalised differential cross sections and therefore only on the relative distribution of the events in the event sample $\{\vec{\cal  X}_1,\ldots, \vec{\cal X}_N\}$.
The total cross section $\sigma(\vec{\alpha})$ and therefore the expected number of observed events may also depend on the model parameters.  To make use of the additional information contained in the size $N$ of the event sample the so-called extended likelihood function can
be applied (see e.g. \Ref{Barlow:1990vc}). Here the number of observed events $N$ is assumed to be a random number distributed according to a Poisson distribution with expectation value
$\nu(\vec{\alpha})=\sigma(\vec{\alpha})L$, where $L$ denotes the integrated luminosity
of the collider. The extended likelihood function is given by the likelihood in \Eq{eq:likelihood} multiplied with the Poisson
probability to observe an event sample of size $N$ under the
parameter-dependent model assumption (here incorporated in
$\sigma(\vec{\alpha})$):
\begin{eqnarray}\label{eq:extLike}
  \nonumber\mathcal{L}_{\text{ext}}\left(\vec{\alpha}\right)
  &=& \frac{\nu(\vec{\alpha})^N}{N!}e^{-\nu(\vec{\alpha})}
  {\cal L}(\vec{\alpha})\\
 &=&  \frac{L^N}{N!}e^{-\sigma(\vec{\alpha})L}
 \prod\limits_{i=1}^{N}\int d\vec{x}\; {d\sigma(\vec{\alpha})\over d\vec{x}}\;
   W(\vec{x},\vec{\cal X}_i).
\end{eqnarray}
In the extended likelihood the normalisation in
front of the differential cross section cancels. The \MEM
employing extended likelihoods is thus not only sensitive to the relative distribution but
also to the total number of observed events.  Using the additional
information contained in the absolute number of observed events may
result in a more efficient estimator when the \MEM is used within the
context of parameter determination.  However, the integrated
luminosity $L$ is needed in \Eq{eq:extLike} which may introduces an
additional uncertainty potentially spoiling the gain
in efficiency. We shall come back to this issue when we discuss as an illustration the
top-quark mass extraction in detail in section \ref{sec:topmassextraction}.
\paragraph{Properties of \MEM estimators}
Since the likelihood as
defined in \Eq{eq:likelihood} depends
on the measured event sample $\{{\cal X}_1,\ldots,{\cal X}_N\}$ which
follows a probability distribution and can thus be seen as a
collection of random
numbers,  the Maximum Likelihood
estimator $\widehat{\vec{\alpha}}(\{{\cal X}_1,\ldots,{\cal X}_N \})$
is itself a random number with expectation value
$E[\widehat{\vec{\alpha}}]$ and variance $V[\widehat{\vec{\alpha}}]$.
The variance of an estimator is bounded from below by the
Rao-Cram\'{e}r-Fr\'echet inequality (see \Ref{cowan1998statistical}).
In the large sample limit Maximum Likelihood estimators approach the
Rao-Cram\'{e}r-Fr\'echet bound making them in a sense maximally efficient. Furthermore,
in this limit Maximum Likelihood estimators
exhibit `asymptotic normality', meaning that they are distributed according
to a Gaussian normal distribution with $\mu=E[\widehat{\vec{\alpha}}]$
and $\sigma^2=V[\widehat{\vec{\alpha}}]$ (see
\Ref{cowan1998statistical}).  A potential disadvantage of parameter extraction based on Maximum Likelihood is that the estimator is prone to have a
non-vanishing bias
\begin{equation}\label{eq:bias}
  \vec{b}=E[\widehat{\vec{\alpha}}]-\vec{\alpha}^{\tiny\mbox{true}},
\end{equation}
if the analysed data is not distributed exactly according to
the assumed probability density. In practice, such biases
can be accounted for by the experiments through a calibration of the method,
however, at the cost of introducing associated uncertainties. Therefore it
is preferable to reduce the bias by modelling the true
probability density as accurately as possible. This is an important
motivation to extend the \MEM to NLO accuracy.

\subsection{The \MEM at next-to-leading order accuracy in QCD}
It is well known that for many processes the Born approximation gives
only a rough estimate and higher-order corrections are sizeable. Furthermore, high energetic hadron scattering leads to an
increase in jet activity which needs to be modelled appropriately
unless the additional jet activity is vetoed which would however lead
to a reduction of the event sample and may also spoil the convergence of the perturbative expansion.
Higher-order corrections lead also to a better modelling of the jets and are mandatory to obtain more realistic predictions. In Born
approximation each parton is identified with a jet. Only beyond
leading order the recombination of two partons to form a jet occurs.
It is thus highly desirable to include higher-order corrections in the
\MEM. When the \MEM is used for parameter determinations higher-order
corrections are crucial: in general, next-to-leading order corrections
are required to unambiguously define the renormalisation scheme. As
long as only leading-order calculations are used, the renormalisation
scheme is not defined. 

Extending the \MEM to include higher-order
corrections is however non-trivial. To understand the origin of the
difficulties we start with the general structure of NLO
corrections. Schematically, the differential cross
section including NLO corrections reads:
\begin{eqnarray}
  \label{eq:NLO-diff-xs}
  {d\sigma^\NLO\over
    dx_1\ldots dx_r} &=& \int dR_n(\vec{y})
  {d\sigma^{BV}\over dy_1\ldots dy_s} {\cal F}^{n}_{J_1,\ldots,J_n}(y_1,\ldots, y_s)
  \delta(\vec{x} - \vec{x}(J_1,\ldots,J_n))\nn\\
  &+& \int dR_{n+1}(\vec{z}) {d\sigma^R\over dz_1\ldots dz_{t}}
  \bigg({\cal F}^{n+1}_{J_1,\ldots,J_n}(z_1,\ldots, z_t)
  \delta(\vec{x} - \vec{x}(J_1,\ldots,J_n))\nn\\
  && +\quad{\cal F}^{n+1}_{J_1,\ldots,J_{n+1}}(z_1,\ldots, z_t)
  \delta(\vec{x} - \vec{x}(J_1,\ldots,J_n,J_{n+1}))\bigg)
\end{eqnarray}
We denote with $d\sigma^{BV}$ ($d\sigma^R$) the Born and virtual (real)
contributions to the next-to-leading order differential cross section.
${\cal F}^{n}_{J_1,\ldots,J_n}$ (${\cal F}^{n+1}_{J_1,\ldots,J_n}$) defines the jet
function which implicitly introduces a mapping of the $n$ $(n+1)$ parton momenta to the $n$ jet
momenta and implements phase space cuts depending on the jet momenta.
${\cal F}^{n+1}_{J_1,\ldots,J_{n+1}}$ defines the jet function for the
$(n+1)$-parton phase space in which all partons are resolved as individual jets and  no recombination occurs.
It is in fact through ${\cal F}^{n+1}_{J_1,\ldots,J_n}$ that a jet is modelled for the first time in a non-trivial way within perturbation
theory. Together
with the contribution due to ${\cal F}^{n+1}_{J_1,\ldots,J_{n+1}}$ this leads to
the aforementioned improved description of additional jet activity. 

As stated above, no recombination can occur in leading order and the jet
momenta are identified with the parton momenta:
\begin{equation}\label{eq:LOmap}
J_i=J_i(p_i)=p_i,\quad i=1,\ldots,n.
\end{equation}
The jet functions for identifying the $n$ partons with $n$ jets have the form
\begin{equation}
{\cal F}^{n}_{J_1,\ldots,J_n}\sim\prod\limits_{i=1}^{m}
\Theta_i(J_1(p_1),\ldots,J_n(p_n))
\end{equation}
with functions $\Theta_i$ which encode whether the $n$ jet momenta are
resolved as a $n$-jet final state.  (The same is also true for the
contribution involving ${\cal F}^{n+1}_{J_1,\ldots,J_{n+1}}$.) 
Since in leading order the mapping is
trivial it has been omitted in \Eq{eq:partonic-diff-xs} although the
jet function might be required to render the cross section finite in
cases in which the leading-order predictions contain already soft
and/or collinear singularities (e.g. $pp\rightarrow t\bar{t}j$). The virtual corrections can thus be treated in a similar way as the Born approximation, ignoring for the moment that they may contain soft and collinear singularities which must be cancelled. In case of the real corrections, the
situation is more involved since now recombination can occur and the
jet momenta are in general non-trivial functions of the partonic
momenta with the mapping depending on the phase space region
\begin{equation}\label{eq:realmap}
  J_i=J_i(p_1,\ldots,p_{n+1}),\quad i=1,\ldots,n.
\end{equation} 
The jet functions for the recombinations of $n+1$ partons to form $n$
jets have the form
\begin{equation}
 {\cal F}^{n+1}_{J_1,\ldots,J_n}\sim\prod
  \limits_{i=1}^{m}\Theta_i(J_1(p_1,\ldots,p_{n+1}),
  \ldots,J_n(p_1,\ldots,p_{n+1}))
\end{equation} 
where the mapping from the $n+1$ parton momenta to the $n$ jet
momenta is given by the functional dependences of the jet momenta on the parton momenta in \Eq{eq:realmap}.
Note that the differential cross section is differential in `jet
variables' and not in partonic variables: We use the function
$\vec{x}(J_1,\ldots,J_n)$ to relate the variables used to describe the
differential cross section to the jet momenta.  Strictly speaking this
detour is not required and one could avoid introducing jets as long as
the definition of variables $\vec{x}$ used to describe the final state
is infrared safe. Using jet momenta in intermediate steps however,
simplifies the construction of infrared safe distributions and leads to a general algorithm to extend the \MEM to NLO accuracy. Inspecting
\Eq{eq:NLO-diff-xs} in more detail various problems, preventing a
straightforward application, become obvious:
\begin{enumerate}
\item The first two contributions are individually
  ill-defined because of soft and collinear singularities. Only in the
  sum of the two a finite result is obtained. 
\item Being differential in the measured jet observables $\vec{x}$ introduces conditions on the recombination of unresolved real parton momenta to form respective jet momenta corresponding to the fixed values in $\vec{x}$. These conditions prevent an
  efficient numerical integration of the real contribution.
\item In the $n$-parton contribution the jets are identified with the
  partons. As a consequence the jet momenta satisfy the $n$-parton
  kinematics, i.e. the jet momenta respect the on-shell condition as
  well as four-momentum conservation. For the recombined jets obtained
  from $n+1$-parton momenta this is in general not the case. In
  particular, defining the momentum $p_j$ of the recombined jet as the sum
  of the combined parton momenta violates the on-shell condition
  unless the recombined momenta are strictly collinear. As a
  consequence the jet momenta cannot be used to evaluate
  $d\sigma^{BV}$ which would be the naive way to establish a point wise
  cancellation of the soft and collinear divergences.\footnote{Note that this is not a problem in conventional parton-level Monte-Carlos which always integrate over finite phase space regions and the cancellation works at the level of the integral.}
\end{enumerate}
In \Ref{Martini:2015fsa} it has been argued, that the last two
obstacles can be overcome using modified jet algorithms relying on a $3\to 2$ clustering instead of a $2\to 1$ recombination usually
applied. The additional spectator particle allows to guarantee
four-momentum conservation and the on-shell conditions at the same
time. In practice, a recombination inspired by the phase space
parameterisation used in the Catani-Seymour dipole subtraction method 
\cite{Catani:1996vz,Catani:2002hc}
is used in \Ref{Martini:2015fsa}. This technique allows a factorisation of the
$n+1$-parton phase space measure into an unresolved and a resolved
contribution. Schematically, this factorisation reads:
\begin{equation}
  dR_{n+1}(\vec{z}) = dR_n(\vec{y}) \times d\Phi, 
\end{equation}
where $dR_n(\vec{y})$ denotes the phase space of the $n$ jets obtained
after recombination, and $d\Phi$ parameterises the unresolved phase
space leading to a clustering of two partons. The phase space measure
$dR_n(\vec{y})$ can thus be identified with the one occurring in the
leading-order calculation. Using this factorisation the second and
third issue are solved. To address the first issue we advocated in
\Ref{Martini:2015fsa} to use the so-called \PSS. For more details we
refer to \Ref{Martini:2015fsa}.  At this point a further remark
concerning the use of the \PSS might be in
order. It is well known that the \PSS is often
numerically challenging and requires significant computing time unless
the phase parametrisation is carefully chosen and optimized to the
specific process. For more involved processes this may lead to severe
restrictions concerning the applicability of the method applied here.
We stress however, that the usage of the \PSS is not
mandatory and could be avoided. In general, subtraction methods may be
used as well, as long as the method itself does not interfere with the
phase space factorisation used here. This excludes the dipole
subtraction method \`a la Catani and Seymour 
(see \Refs{Catani:1996vz,Catani:2002hc}) as has been pointed out in \Ref{Martini:2015fsa}. However, the
Frixione-Kunszt-Signer  type subtraction 
(see \Refs{Frixione:1995ms,Frixione:1997np}) should be applicable. This
shall be studied in the future.  

In the following section we provide some further details on the
modified jet algorithm and the phase space parameterisation used in
this article.  In \Ref{Baumeister:2016maz} an alternative method to
circumvent the two last issues has been presented to avoid introducing
modified jet algorithms. However, no implementation of the method has
been presented so far and it remains to be shown that the practical
application of this approach is not limited by the required computer
resources.  For the purpose of this article we stick to the method
presented in \Ref{Martini:2015fsa} to explore the potential of the
\MEM extended to NLO accuracy and leave the implementation of the
method presented in \Ref{Baumeister:2016maz} for future studies.

\subsection{Modified jet algorithm and phase space parameterisation}
\label{sec:jetalg}
As mentioned before we augment 
an existing $2\to 1$ algorithm by using a modified 
$3\to 2$ clustering.
As an example we study the $k_t$-jet algorithm for hadron colliders with the resolution
criteria defined by (see \Ref{Catani:1993hr})
\begin{equation}
  d_{ij}=\min\left({p^{\perp}_{i}}^2,{p^{\perp}_{j}}^2\right)
  \frac{\Delta R^2_{ij}}{R^2},\quad 
  \Delta R^2_{ij}=\left(y_i-y_j\right)^2+\left(\phi_i-\phi_j\right)^2,
  \quad 
  d_{iB}={p^{\perp}_{i}}^2.
\end{equation}
Based on these resolution criteria we define a $3\to 2$
clustering algorithm as follows:
\begin{enumerate}
\item Pick final state partons $i$ and $j$ or final state parton
  $i$ and beam $B$ with minimal $d_{ij}$ or
  $d_{iB}$ to be clustered.
\item 
  \begin{enumerate}
  \item  If $d_{ij}$ is minimal, pick spectator
    parton from final state ($k$) or beam ($a$).
  \item If $d_{iB}$ is minimal, pick beam particle
  ($a$) and spectator parton from final state ($k$) or beam ($b$).
  \end{enumerate}
\item Cluster ($ijk$), ($ija$), ($iak$) or ($iab$) according to the
  respective $3\to 2$ phase space mapping introduced in the 
  Catani-Seymour subtraction method for the dipoles
  ${\cal D}_{ijk}, {\cal D}_{ija}, {\cal D}_{iak}, {\cal D}_{iab}$ 
  \cite{Catani:1996vz,Catani:2002hc}.
\end{enumerate}
After clustering, only the jets that pass the experimental cuts 
\begin{equation}\label{eq:expcuts}
p^\perp>p^\perp_{\text{min}},\quad |\eta|<\eta_{\text{max}}
\end{equation}
are kept in the list of resolved jets.
In contrast to the widely used anti-$k_t$-jet algorithm (see \Ref{Cacciari:2008gp}) the $k_t$-jet algorithm can be formulated in an `exclusive variant' where exactly a desired number of jets is required to be resolved which is strongly discouraged for the anti-$k_t$-jet algorithm\footnote{Ref~\cite[p. 21]{Cacciari:2011ma}: ``We advise against the use of exclusive jets
in the context of the anti-$k_ t$ algorithm, because of the lack of physically meaningful hierarchy in the
clustering sequence.''} . For example, we define the `exclusive single top-quark production' by requiring a resolved top-tagged jet\footnote{For simplicity we do not consider the decay of the top quark but treat it as being resolved as a top-tagged jet. Including also the top-quark decay does not pose any principle challenge to the presented methodology.} accompanied by exactly one light jet
\begin{equation}\label{eq:excldef}
p+p\rightarrow t+j.
\end{equation}
We stress that by requiring the signal signature in \Eq{eq:excldef} we focus on the part of the NLO corrections which leads to the same signal signature as the Born process.
On the contrary, in the `inclusive variant' of the jet algorithm also events with several jet multiplicities are included.
Correspondingly, we define the `inclusive single top-quark production' by requiring a resolved top-tagged jet accompanied by at least one light jet
\begin{equation}\label{eq:incldef}
p+p\rightarrow t+j\;(+X).
\end{equation}
Contrariwise to the exclusive case the signal signature in \Eq{eq:incldef} allows to include the real contribution corresponding to an additional resolved jet $X$ in the NLO corrections.

Compared to the usual definition of sequential $2\to 1$ jet algorithms (cf. e.g. \Ref{Salam:2009jx}) only step three is modified in the $3\to 2$ jet algorithm used here. Instead
of the four-momentum sum $p_i+p_j$ the partons are clustered using the $3\to 2$ mapping introduced within the dipole subtraction method to form the jet momentum $J_j(p_i,p_j,p_k)$. Note that
the clustering offers the additional freedom to choose a spectator
particle. In principle, it is possible to always choose the same
spectator particle (as long as the particle itself is not collinear or
soft). However, making different choices for the spectator particle,
the additional freedom can be used to reduce the difference of the
modified clustering prescription with respect to the conventional
$2\to 1$ recombination $p_i+p_j$. 
In order to quantify the difference between the two schemes we study the difference between the respective clustered jet momenta for a recombination of two unresolved final-state partons $i$ and $j$   
\begin{equation}\label{eq:fnorm}
||J_j(p_i,p_j,p_k)-(p_i+p_j)||=
    \max\left(\left|J^0_j(p_i,p_j,p_k)-(p^0_i+p^0_j)\right|,
    \left|\vec{J}_j(p_i,p_j,p_k)-(\vec{p}_i+\vec{p}_j)\right|\right).
\end{equation}
It is thus possible to try different final- or initial-state particles as spectator $k$ and choose
the one which minimises the quantity given in the right-hand side of \Eq{eq:fnorm}.  
In case of unresolved radiation $i$ which is associated with the beam, we have the freedom not only to choose  a final- or initial-state spectator $k$ but also the beam particle $a$ to be clustered with. In $2\to 1$ clustering prescriptions additional radiation too close to the beam is usually just omitted from the event without altering the other final state momenta. We can quantify the difference between the two schemes by studying the difference in the resulting final states
\begin{equation}\label{eq:inorm}
  ||\sum\limits_m {J}_m(p_i,p_a,p_k) - \sum\limits_{m\neq i} p_m||=\max\left(\left|\sum\limits_m {J}^0_m(p_i,p_a,p_k) - \sum\limits_{m\neq i} p^0_m\right|,\left|\sum\limits_m \vec{J}_m(p_i,p_a,p_k) - \sum\limits_{m\neq i} \vec{p}_m\right|\right).
\end{equation}
For given unresolved radiation $i$ we can choose the beam particle $a$ together with the initial- or final-state spectator $k$ to minimise the quantity given in the right-hand side of \Eq{eq:inorm}. We stress that regarding the computing time of the jet algorithm, the additional loop over all possible spectator particles in \Eq{eq:fnorm} and \Eq{eq:inorm} could be avoided by sticking to simpler criteria for choosing the spectator. As noted above, the choice of the spectator is arbitrary and one could also choose the same (non-soft/non-collinear) particle (e.g. from the beam). The actual choice of the spectator has no influence on the validity of the presented algorithm. We use the specific setup to minimise the effects with respect to commonly used $2\to1$-algorithms.

While the modified clustering is mainly
introduced to address the aforementioned problems it has been argued
in \Ref{Martini:2015fsa} that it might also provide a cleaner separation of
perturbative and non-perturbative effects. The jet mass produced
applying the momentum sum has in general very little to do with the
observed jet masses which are mostly due to non-perturbative effects. 
Since the modified clustering is a crucial
part we shall discuss it in more detail. As a concrete example we
illustrate in the following the case that two final-state partons $i$ and $j$ are clustered with a final-state spectator $k$.

As a
technical detail we add that we have used in the current analysis a
slightly different parameterisation of the unresolved phase space
measure compared to \Ref{Martini:2015fsa}. This leads to an improved
convergence of the numerical integration over the unresolved phase space in the
context of the so-called \PSS
\cite{Giele:1991vf,Giele:1993dj} which we adopt to regularise soft and
collinear divergences. The details are given in appendix
\ref{sec:ps-parameterisation}.
In case the spectator $k$ (with mass $m_k$) is chosen from the final
state we may factorise the phase space measure using the
parameterisation as described in detail in
\Refs{Catani:1996vz,Catani:2002hc} within the context of the dipole
subtraction method.  For example, for two massive partons $i$ and $j$
with masses $m_i$ and $m_j$ the parameterisation reads
\cite[(5.11)]{Catani:2002hc}:
\begin{equation}\label{eq:psfacexffm}
   dR_{n+1}\left(P,p_1,\ldots,p_{n-2},p_i,p_k,p_{j}\right)
  = dR_{n}\left(P,J_1,\ldots,J_{n-2},J_j,J_{k}\right)dR_{ij,k}(J_j,J_k,\vec{\Phi})
\end{equation}
where $\vec{\Phi}$ denotes the collection of variables used to parameterise the
`unresolved' phase space. As unresolved regions we define the phase space
regions in which two partons cannot be resolved any longer (according
to our jet resolution criteria) as separate partons and are thus
clustered to form a single jet.
The  $n$-body phase space measure is
defined as usual by:
\begin{equation}
  \label{eq:phasespace}
   dR_{n}(P,p_1,\ldots,p_n) = (2\pi)^4\delta(P - \sum_i p_i)
   \prod_{i=1}^n {d^4p_i\over (2\pi)^3} \delta_+(p_i^2-m_i^2).
\end{equation}
The mapping 
\begin{equation}
  p_1,\ldots,p_n,p_{n+1} \to J_1,\ldots,J_n,\vec{\Phi},
\end{equation}
induces a clustering prescription
\begin{equation}
  p_i,p_j,p_k \to J_{j}, J_k 
\end{equation}
($p_k$ and $J_k$ denote the momenta of the spectator before and after
the clustering). For \Eq{eq:psfacexffm} the mapping is given explicitly in the appendix (see \Eq{eq:ffclus}).
As has been shown in \Ref{Martini:2015fsa} the mapping is invertible.
Given the jet momenta $J_1,\ldots,J_n$ it is thus straightforward to
identify the unresolved phase space regions leading to the recombined
jets $J_1,\ldots,J_n$ (see also appendix \ref{sec:ps-parameterisation}). Identifying in \Eq{eq:psfacexffm} the
phase space of the $n$ recombined jets as the $n$-body phase space used in the Born and virtual contribution in \Eq{eq:NLO-diff-xs}, the integration of the unresolved contributions of the real corrections can be simplified allowing a direct integration:
\begin{eqnarray}
\label{eq:phase-space-factorisation}
 && \int dR_{n+1}(\vec{z}) {d\sigma^R\over dz_1\ldots dz_{t}}
  \,{\cal F}^{n+1}_{J_1,\ldots,J_n}(z_1,\ldots, z_t)
  \delta(\vec{x} - \vec{x}(J_1,\ldots,J_n))\nn\\
  &=& \int dR_{n}(\vec{y}) {\cal F}^{n}_{J_1,\ldots,J_n}(y_1,\ldots, y_s)
  \delta(\vec{x} - \vec{x}(J_1,\ldots,J_n))
  \int dR_{ij,k} {d\sigma^R\over dz_1\ldots dz_{t}} 
  .
\end{eqnarray}
The variables $z_1,\ldots,z_t$ are expressed in terms of the jet
momenta $J_1,\ldots,J_n$ (or equivalently the variables
$\vec{y}=(y_1,\ldots,y_s)$) and the variables in $\vec{\Phi}$ used to
parameterise the unresolved phase space. We may define the last part
of \Eq{eq:phase-space-factorisation} as the real part of a
differential jet cross section at NLO accuracy
\begin{equation}
   {d\sigma^R\over dR_n(J_1,\ldots,J_n)}=  \int dR_{ij,k} {d\sigma^R\over dz_1\ldots dz_{t}}.
\end{equation}
Combining this contribution (a similar factorisation holds for all
unresolved contributions) with the virtual corrections and the Born
contribution allows us to define a differential jet cross section at
NLO accuracy (cf. \Eq{eq:NLO-diff-xs})
 \begin{equation}  \label{eq:jet-xs}
   {d\sigma^\NLO\over dR_n(J_1,\ldots,J_n)}= {d\sigma^{BV}\over dR_n(J_1,\ldots,J_n)}+{d\sigma^R\over dR_n(J_1,\ldots,J_n)}.
\end{equation}
Note that the formalism presented here allows as an important
application the generation of unweighted jet events following the NLO
cross section.

To end this section let us add a comment on the use of non-standard
jet algorithms. Although the modified jet clustering as advocated here
may have some theoretical advantages, for example potentially
providing a clearer separation of perturbative and non-perturbative
aspects, a significant amount of work is required from the
experimental side before the modified algorithms can be used in
praxis. Tuning as well as studies of experimental uncertainties need
to be redone using the modified setting. Obviously, this effort is
only justified if an increased precision can be reached in the end.
One purpose of this article is therefore to explore the reachable
accuracy. We also note---as will be shown in a future
publication---that by restricting the kinematic variables used in the
analysis, the modification of the jet algorithm can be avoided.

\section{Event generation with NLO event weights}
\label{sec:unweighted-events} 
In this section we use the differential jet cross section from
\Eq{eq:jet-xs} as an event weight to generate unweighted jet events
which are distributed strictly according to the fixed-order NLO cross section. Note that the possibility to generate unweighted jet-events following the NLO cross section is a substantial progress compared to established methods. As concrete
example processes we study the $s$- and $t$-channel single top-quark
production at the LHC. After a brief review of the calculational setup
we define exclusive (cf. \Eq{eq:excldef}) and inclusive events (cf.
\Eq{eq:incldef}) by specifying jet observables to characterise the
events. The construction (and implementation) of the differential jet
cross section (\Eq{eq:jet-xs}) including full NLO effects is a
non-trivial, error-prone task. Hence, we start by thoroughly
scrutinising the phase space factorisations entering \Eq{eq:jet-xs} as
well as the consistency of the \PSS used to regulate the IR
divergences of the differential NLO calculation. We then show that the
generated events are indeed distributed according to the NLO cross
section.

\subsection{Preliminaries}
\subsubsection{Setup}
For this work the NLO corrections for single top $s$- and $t$-channel
production employing the \PSS to subtract
the infrared divergences are taken from \Ref{Cao:2004ky}.

If not stated otherwise, we use the following setting throughout this
work: All calculations are done for proton-proton collisions at the
LHC with a centre-of-mass energy of $\sqrt{s}=13$ TeV. For the parton
distribution functions (PDFs) we use the PDF set 
`MSTW2008nnlo68cl' \cite{Martin:2009iq}. To be consistent with the PDF
evolution the QCD coupling constant $\alpha_s$ is taken
from the PDF set. The electromagnetic
fine structure constant is set to
\begin{equation}
  \alpha(m_Z)=1/132.2332298. 
\end{equation}
Note that $\alpha(m_Z)$ appears as an overall factor which cancels in 
cross section ratios or can be easily adjusted to other values in the
cross section predictions.
For the masses of the electroweak gauge bosons we use
\begin{equation}
  m_Z=91.1876\ \GeV\quad \mbox{and}\quad  m_W=80.385 \mbox{ GeV}.
\end{equation}
For the weak mixing angle we use the on-shell value defined through
\begin{equation}
  \sin^2(\theta_w) = 1 - {m_W^2\over m_Z^2}.
\end{equation}
The top-quark mass renormalised in the pole-mass scheme is set to
\begin{equation}
  m_t=173.2\ \GeV.
\end{equation}
For the renormalisation (factorisation) scale \muR (\muF) we choose
a dynamical scale. As central scale choice we set
\begin{equation}
\label{eq:scale-definition-excl}
\mu_R=\mu_F=\mu_0=\sum\limits_{m}E^{\perp}_m
\end{equation} 
where the transverse energy is defined as $E^{\perp}=E \sin(\theta)$. The sum in \Eq{eq:scale-definition-excl} runs over all resolved final state jets (cf. \Eq{eq:excldef} or \Eq{eq:incldef}).
The jets are defined according to the jet algorithm as described in 
section \ref{sec:jetalg} (with $R=1$ as a common setting for the exclusive formulation (see e.g. \Ref{Cacciari:2011ma})).
For the experimental cuts on resolved final-state objects we set (cf. \Eq{eq:expcuts})
\begin{equation}
p^{\perp}_{\text{min}}=30\ \GeV,\quad \eta_{\text{max}}=3.5.
\end{equation}
We assume that the detector is blind outside these cuts.

\subsubsection{\PSSP dependence}
\begin{figure}[htbp]
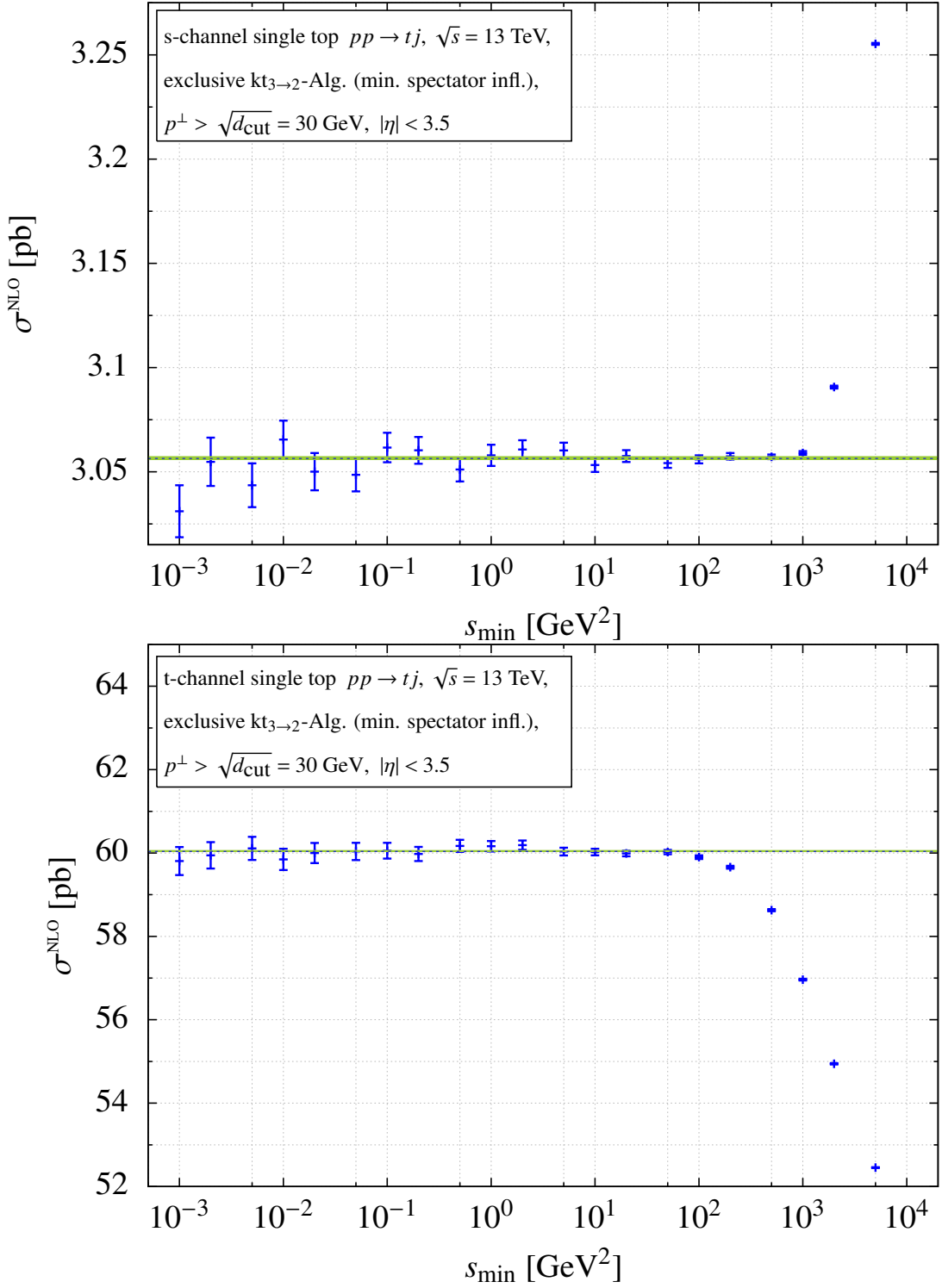

  \begin{flushright}
    \leavevmode
    \includegraphics[height=0.47\textheight]{{{%
          sgtsKT3-2ycut30sminc-50x1e7-crop}}}
    \includegraphics[height=0.47\textheight]{{{%
          sgttKT3-2ycut30sminc-50x1e7-crop}}}
     \caption{Fiducial NLO cross section for the $s$-channel (upper plot) and $t$-channel (lower plot)
       production of a top-tagged and a light jet as a function of the
       \PSSP \smin.}
    \label{fig:sminc_sgt}
  \end{flushright}
\end{figure}    
\begin{figure}[htbp]
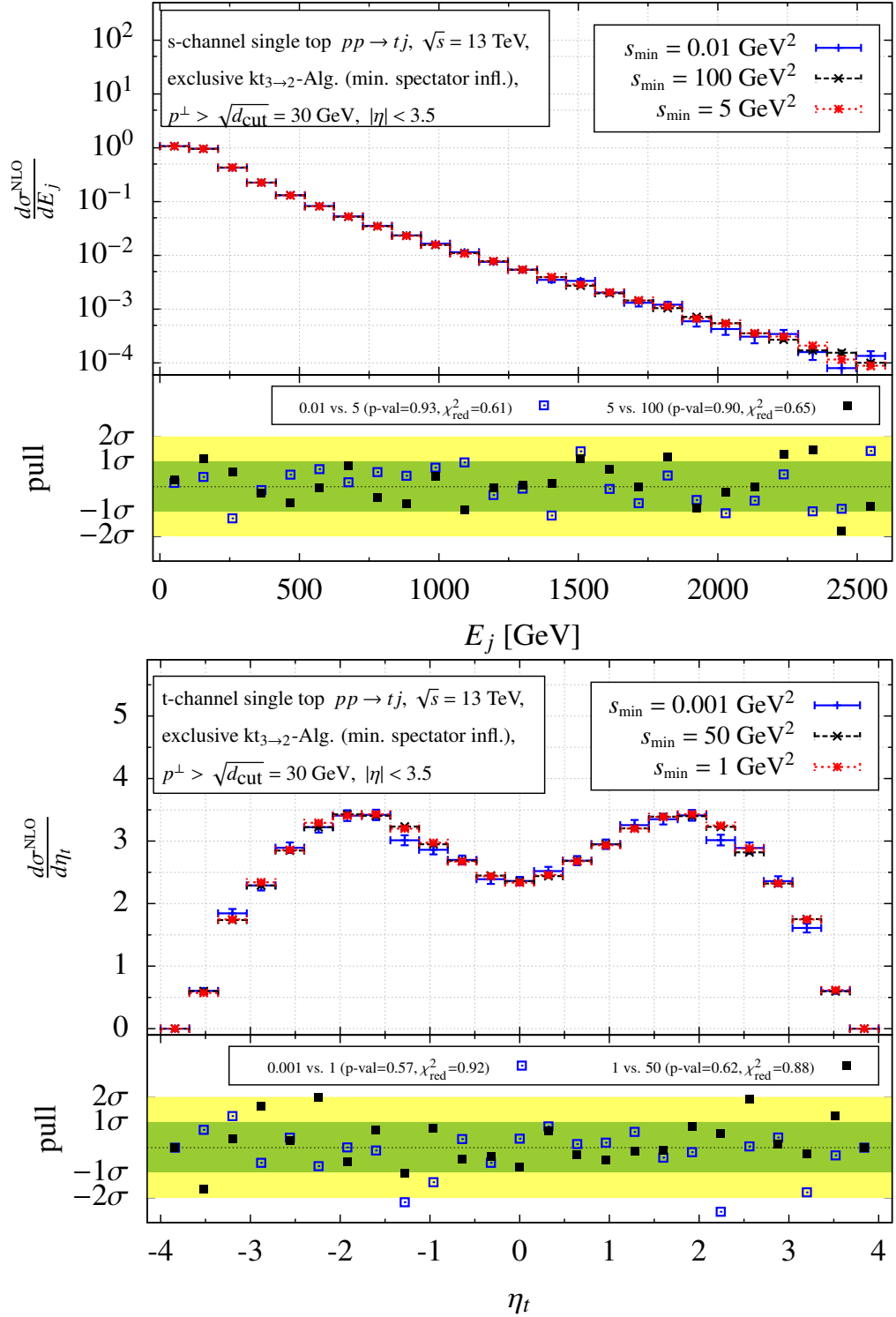

  \begin{flushright}
    \leavevmode
    \includegraphics[height=0.47\textheight]{{{%
          sgtsKT3-2ycut30compmnspecE2-50x1e7sminc-crop}}}
    \includegraphics[height=0.47\textheight]{{{%
          sgttKT3-2ycut30compmnspecETA1-50x1e7sminc-crop}}}
    \caption{Energy distribution of the light jet 
      from $s$-channel single top production (upper plot) and pseudo rapidity distribution of the top-tagged jet from $t$-channel single top production (lower plot) for three different values
      of the slicing parameter \smin.}
    \label{fig:sminc_sgtst}
  \end{flushright}
\end{figure}
Virtual and real contributions in \Eq{eq:NLO-diff-xs} are individually
IR divergent. Only their sum is finite for infrared-safe observables
as guaranteed by the Kinoshita-Lee-Nauenberg theorem. In practical
calculations, we thus need to regularise the related divergences and
cancel them when combining real and virtual corrections.  Following
\Ref{Cao:2004ky} we employ the \PSS (see also
\Refs{Giele:1991vf,Giele:1993dj}) to achieve this cancellation.
Within the \PSS the phase space for the real corrections is split into
regions where the squared matrix elements contains soft and collinear
divergences and the remaining phase space. The latter can be
integrated numerically in $4$ dimensions. Using soft and collinear
approximations the unresolved contributions can be analytically
integrated in a process-independent manner and the singularities can
be cancelled with the ones in the virtual corrections.  Note that the
separation of the phase space into `singular' regions and `finite'
regions should not interfere with the jet clustering. We employ the
so-called one cut-off \PSS method in the form as presented in
\Ref{Cao:2004ky} with the slicing parameter $\smin$, controlling the
separation into singular and finite phase space regions. Since the
slicing is to some extent arbitrary the result must be independent of
$\smin$.  However, because of the soft and collinear approximations
applied in the singular regions a systematic error is introduced and
the result is no longer independent of the slicing parameter. Since
the error scales with \smin it can be neglected for sufficiently small
\smin.  On the other hand, since singular regions and finite regions
both individually depend logarithmically on \smin---only in the sum we
find the aforementioned linear dependence---the statistical error of
the numerical Monte-Carlo integration grows with smaller \smin and it
takes more time to perform the numerical integration. An important
test within the \PSS is thus to numerically establish the approximate
independence of the final results of the choice of \smin for
sufficiently small values and to find a compromise between statistical
and systematic uncertainties. 

Because $s$-channel and $t$-channel largely differ in size, we perform
this analysis for each channel separately since otherwise potential
problems in the $s$-channel could hide because of the large
$t$-channel contribution.  The fact that the $s$-channel gives only a
small contribution to single top-quark production which is difficult
to establish experimentally makes the application of the \MEM to the
$s$-channel also particularly interesting. The \MEM might help in separating
the $s$-channel contribution from $t$-channel production. 

\FIG{fig:sminc_sgt} shows the fiducial cross sections for the
$s$-channel and $t$-channel production of single top quarks in association with a
light jet at NLO accuracy as a function of the slicing parameter
$\smin$.  For $\smin<1000\ \GeV^2$ ($s$-channel) and $\smin<100\ \GeV^2$ ($t$-channel) \Fig{fig:sminc_sgt} shows plateaus of the fiducial cross sections compatible with constant values
within the statistical errors. The straight dashed lines show the
results of fits to the first $17$ ($s$-channel) and $13$ ($t$-channel) data points assuming constant
cross sections.  The fits yield $\sigma^\NLO=3.0564\pm0.0008$ pb for the $s$-channel and $\sigma^\NLO=60.04\pm0.02$ pb for the $t$-channel.
\FIG{fig:sminc_sgt} also illustrates the increasing statistical
uncertainties for decreasing \smin values.  The \smin dependence as
shown in \Fig{fig:sminc_sgt} suggests to choose a value for the
slicing parameter well below $1000\ \GeV^2$ for the $s$-channel and $100\ \GeV^2$ for the $t$-channel.  As a compromise between
statistical and systematic uncertainties we choose $\smin=5\ \GeV^2$ for the $s$-channel and $\smin=1\ \GeV^2$ for the $t$-channel.
Concerning the total cross sections this seems rather small given the
results shown in \Fig{fig:sminc_sgt}.
However, differential distributions which we study next, may
introduce additional scales and may be more sensitive to \smin. In
this context $5\ \GeV^2$ and $1\ \GeV^2$ seem to be a good choice since they should be well
below all relevant physical scales and are still large enough to
prevent large statistical fluctuations in the numerical integration.
\FIG{fig:sminc_sgtst} shows distributions of the top-tagged jet and the light jet for
three different values of \smin. In the plots we show results obtained
with the nominal values $\smin=5\ \GeV^2$ ($s$-channel) and $\smin=1\ \GeV^2$ ($t$-channel) as discussed above and results
obtained for lower values of $\smin=0.01\ \GeV^2$ (respectively $\smin=0.001\ \GeV^2$) and higher values of $\smin=100\ \GeV^2$  (respectively $\smin=50\ \GeV^2$).  At the
bottom of each plot the difference of the results obtained with
different \smin are shown normalised to the statistical uncertainty
(`pull').  As we can see from \Fig{fig:sminc_sgtst} (and various other distributions which we also checked but do not show here) any \smin dependence of
the differential distributions is at most of the order of the
statistical uncertainties and thus negligible. While this conclusion
is already obvious from the distribution of the pull in the lower
plots, we also calculated, as a more quantitative measure, the
corresponding $p$-value and the reduced $\chi^2$ for the comparison of
the histograms as described in \Ref{Gagunashvili:2007zz} and
implemented in \Refs{Brun:1997pa,Antcheva:2009zz,Moneta:2008zza}.
From the results we conclude that within the uncertainties the
distributions obtained for different \smin agree with each other.  The
chosen \smin values are thus sufficiently small also for differential
cross sections. Note that for the
aforementioned studies we used the factorised phase space
parameterisation as described in section \ref{sec:jetalg}.  Since this
involves different parameterisations in different (singular) phase space regions,
the approximate slicing-parameter independence of the differential distributions also serves as a first consistency check of their implementation.

\subsubsection{Validation of the phase space parameterisation}
\begin{figure}[htbp]
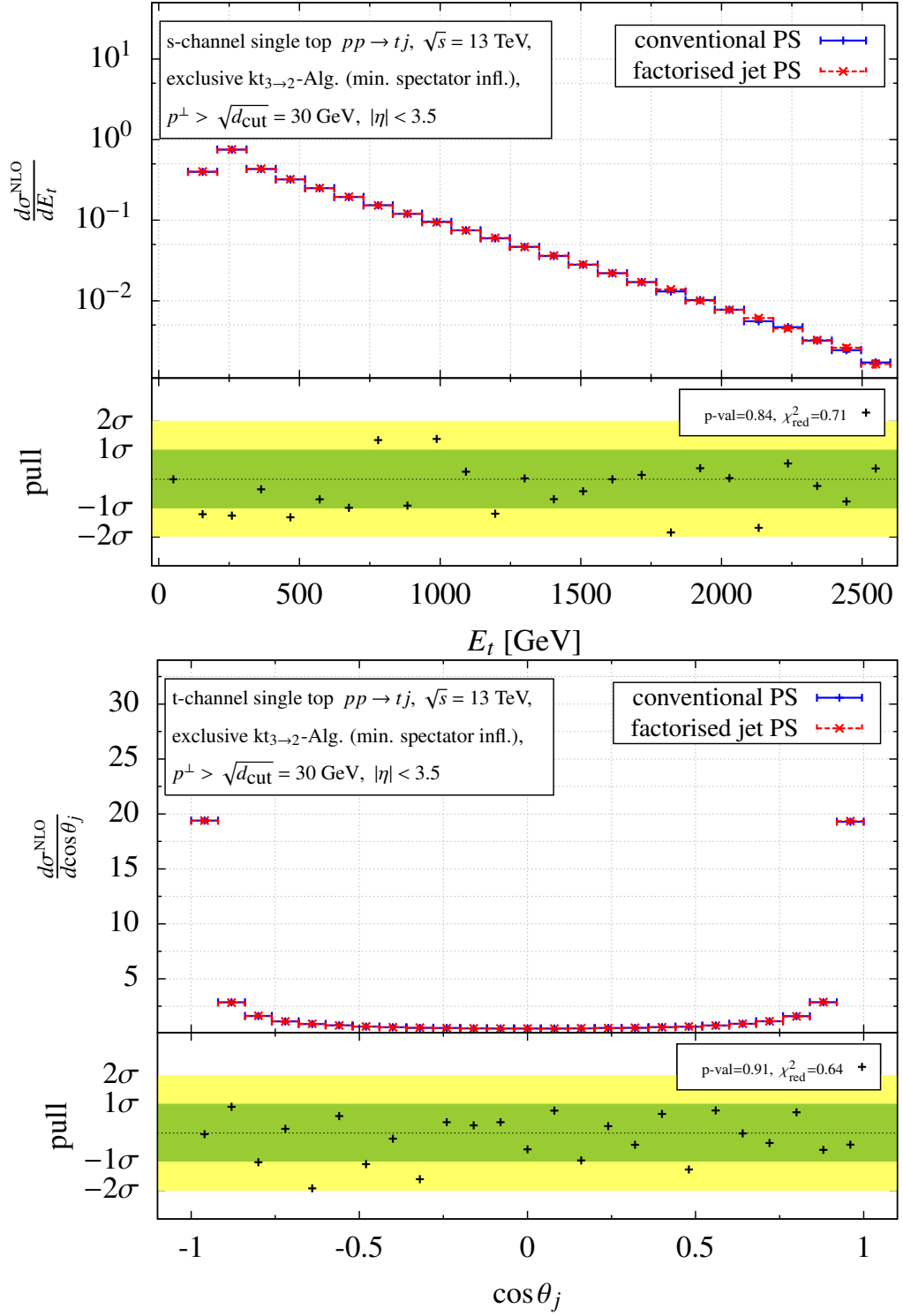

  \begin{flushright}
    \leavevmode
    \includegraphics[height=0.47\textheight]{{{%
          sgtsKT3-2ycut30compmnspecE1-50x1e7-crop}}}
    \includegraphics[height=0.47\textheight]{{{%
          sgttKT3-2ycut30compmnspecCTH2-50x1e7-crop}}}  
    \caption{Energy distribution of the top-tagged jet from the $s$-channel (upper plot) and the distribution of the polar angle of the light jet from the $t$-channel (lower plot)
      calculated using a conventional parton level MC (blue, solid) compared to a
      calculation using the factorised jet phase space (red, dashed) as described in
      section \ref{sec:jetalg}.}
    \label{fig:partonMCvsjetMC_sgtst}
  \end{flushright}
\end{figure}
As mentioned above depending on which partons are clustered and how
the spectator is chosen different phase space parameterisations are
used in different phase space regions in case of the real corrections. The implementation involves some
combinatorics and the coding of the formulas as presented in
the appendix \ref{sec:ps-parameterisation} and in \Ref{Martini:2015fsa}.
To validate the phase space parameterisation we reproduce
various distributions of $n$-jet observables at NLO accuracy that have been
calculated with a conventional parton level Monte-Carlo generator in
combination with the $3\to 2$ jet algorithm. Using the
factorised phase space parameterisation any distribution
of an $n$-jet observable
$O\left(\widetilde{J}_{1},\dots,\widetilde{J}_{n}\right)$  can
also be calculated at NLO by integrating the differential jet cross section
as defined in \Eq{eq:jet-xs} over the $n$ jet momenta:
\begin{equation}\label{eq:diffobs}
\frac{d\sigma^{\text{NLO}}}{dO\left(\widetilde{J}_{1},
    \dots,\widetilde{J}_{n}\right)}
  =\int dR_n(J_1,\ldots,J_n)
  \frac{d\sigma^{\text{NLO}}}{dR_n(J_1,\ldots,J_n)}
  \delta\left(O\left(J_{1},\dots,J_{n}\right)
    -O\left(\widetilde{J}_{1},\dots,\widetilde{J}_{n}\right)\right).
\end{equation}
Comparing the two approaches allows us to check the implementation of the phase
parameterisations. In addition, the identification of the $n$ jet momenta given by the $n$-body phase space in the factorised phase space in \Eq{eq:diffobs} with the momenta of the jets obtained from the corresponding $3\to 2$ jet algorithm is checked.
As examples, we study the energy distribution of the top-tagged jet for the $s$-channel and the polar angle distribution of the light jet for the $t$-channel in \Fig{fig:partonMCvsjetMC_sgtst}.
The distributions calculated using the conventional parton level Monte-Carlo generator employing the $3\to 2$ jet clustering are shown as blue
lines while the ones obtained according to \Eq{eq:diffobs} are shown as
dashed red lines. Their difference normalised to the statistical
uncertainty is shown at the bottom of each plot.  In addition, we show
again $p$-value and normalised $\chi^2$ as introduced in the previous
section.  Since the distributions have not been normalised the cross
check of the differential cross sections also serves as a validation of the fiducial cross sections.  We study the exclusive case since we are especially interested in the
case when real radiation is unresolved and there is a non-trivial
mapping from partonic to jet momenta.
(The contribution from events with an additional jet would be
calculated according to the last line of \Eq{eq:NLO-diff-xs} and does
not involve any technical or conceptual problems.)

Inspecting the plots in \Fig{fig:partonMCvsjetMC_sgtst} (and various other distributions which we also checked but do not show here)
we find perfect agreement within the statistical uncertainties between the
two approaches. The results for the $p$-values and the reduced $\chi^2$ further support
this interpretation.

\subsection{Generation of unweighted single top-quark events distributed according to the NLO cross section}
\label{sec:event-generation}
Interpreting the differential jet cross section as an event weight
allows to use \Eq{eq:jet-xs} to generate unweighted jet events following the NLO
cross sections. As described in \Ref{Martini:2015fsa} we use a simple
Acceptance-Rejection algorithm to unweight the NLO events.  To validate
this procedure, we first generate unweighted events and recalculate distributions of $n$-jet observables
that have been calculated in the previous section using the conventional
parton level Monte-Carlo. 

\subsubsection{Exclusive event definition}
\label{sec:event-definition-exclusive-case}
Having in mind the extraction of the top-quark mass with the \MEM from our generated events, we have to specify an event definition that does
not make any assumption on the top-quark mass. For example, one may
use energies and angles to define the measured events in
our pseudo experiment.  Again, we use the $3\to 2$
clustering prescription introduced in the previous section. Because in
current experiments a $2\to 1$ recombination is used, we
study first the impact of the modified recombination procedure. Since the difference in the results using different recombination procedures is an NLO effect, this
study provides also valuable information to minimise the impact of NLO corrections by choosing a sensible event
definition. We study first the exclusive case where we demand precisely
one additional light jet and veto events with more than one light jet.
\paragraph{Impact of the new jet clustering}
\begin{figure}[htbp]
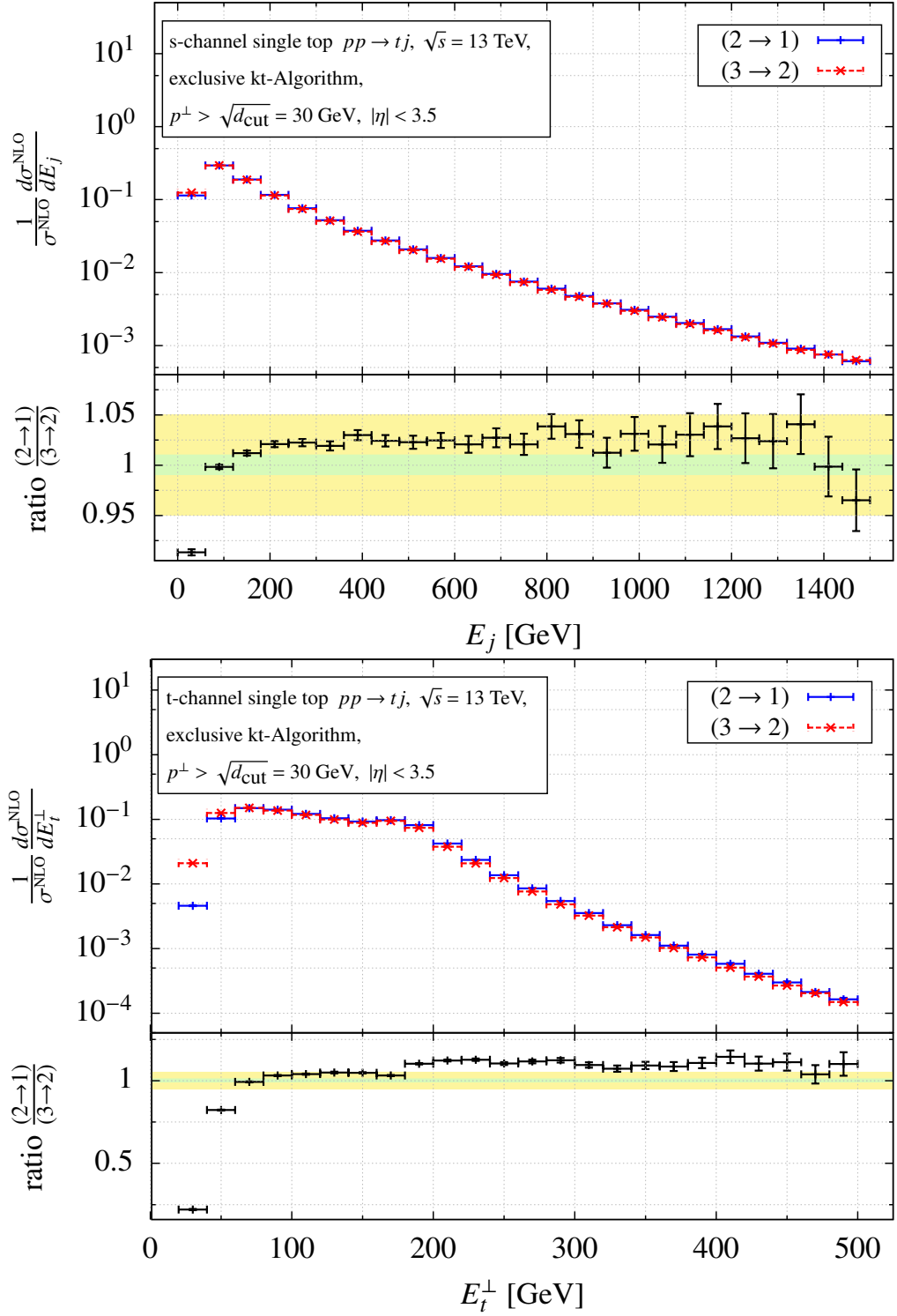

  \begin{flushright}
    \leavevmode
    \includegraphics[height=0.47\textheight]{{{%
          sgtsKT3-2ycut30compjetE2-200x1e7-crop}}}
    \includegraphics[height=0.47\textheight]{{{%
          sgttKT3-2ycut30compjetET1-200x1e7-crop}}}  
    \caption{Examples for (transverse) energy distributions for the top-tagged and the light jet 
      from $s$- and $t$-channel single top-quark production at NLO with
      $3\to 2$ (red,  dashed) and $2\to 1$ (blue, solid) jet clusterings.}
    \label{fig:jetcom_sgten}
  \end{flushright}
\end{figure}
\begin{figure}[htbp]
  \begin{flushright}
    \leavevmode
    \includegraphics[height=0.47\textheight]{{{%
          sgtsKT3-2ycut30compjetETA1-200x1e7-crop}}}
    \includegraphics[height=0.47\textheight]{{{%
          sgttKT3-2ycut30compjetETA2-200x1e7-crop}}}  
    \caption{Examples for pseudo rapidity distributions for the top-tagged and the light jet 
      from $s$- and $t$-channel single top-quark production at NLO with
      $3\to 2$ (red,  dashed) and $2\to 1$ (blue, solid) jet clusterings.}
    \label{fig:jetcom_sgtan}
  \end{flushright}
\end{figure}
\begin{figure}[htbp]
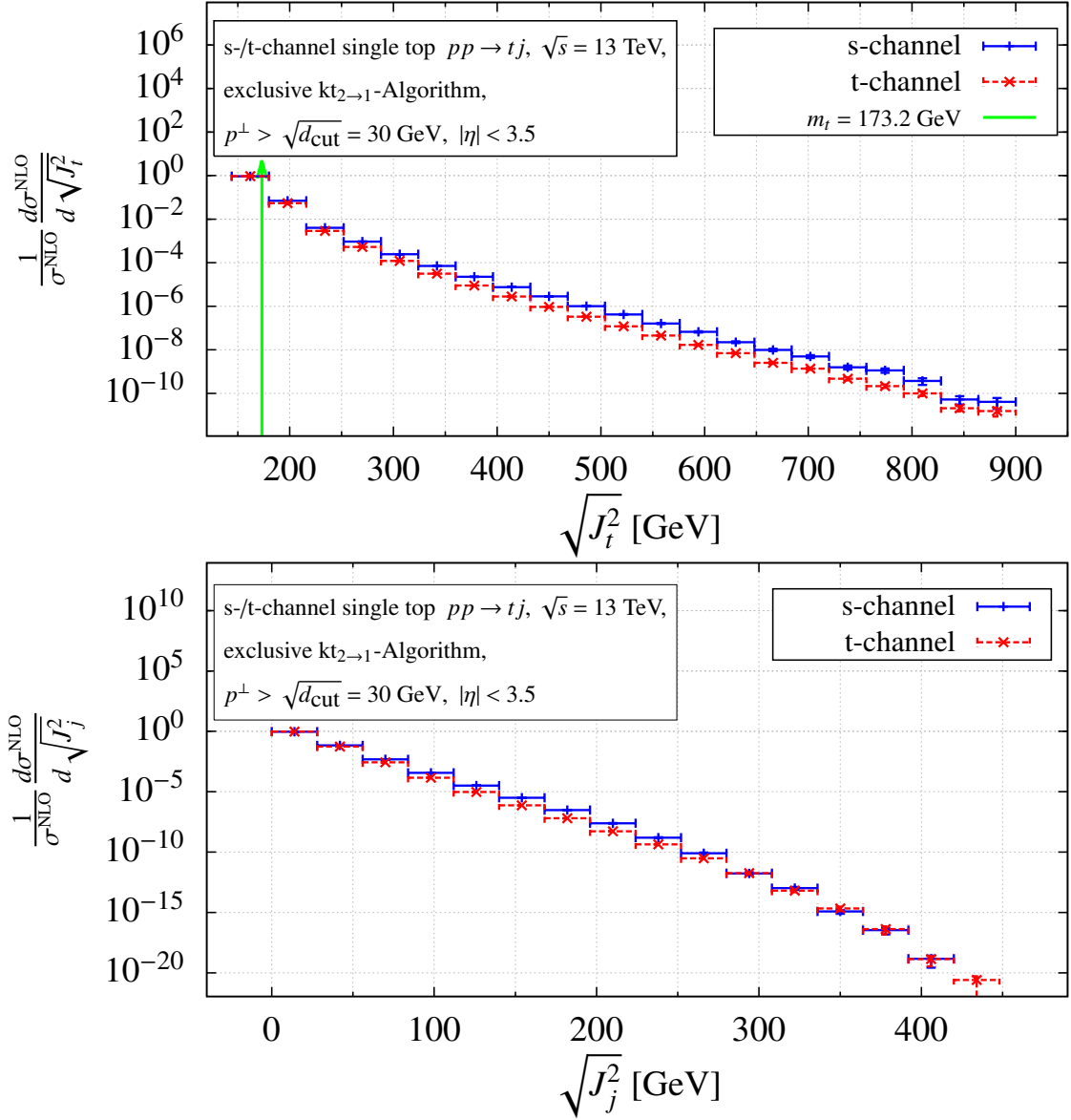

  \begin{flushright}
    \leavevmode
    \includegraphics[height=0.37\textheight]{{{%
          sgtKT3-2ycut30compjetM1-200x1e7-crop}}}  
    \includegraphics[height=0.37\textheight]{{{%
          sgtKT3-2ycut30compjetM2-200x1e7-crop}}}
    \caption{Distribution of the jet masses for $2\to 1$ 
      jet clusterings.}
    \label{fig:jetcom_sgtm}
  \end{flushright}
\end{figure}
The \MEM at NLO, as described in \Ref{Martini:2015fsa}, is only
sensitive to normalised distributions (shapes) but not to the total number
of measured events. Obviously, the distributions are affected by the
jet algorithm. In \Fig{fig:jetcom_sgten} and \Fig{fig:jetcom_sgtan} (and also \Fig{fig:jetcom_sgten1app} to \Fig{fig:jetcom_sgtpr2app} in appendix~\ref{sec:comparison-clustering}) we
illustrate the impact of the modified $3\to 2$ clustering
prescription as introduced in \Ref{Martini:2015fsa} and reviewed in
the previous section for different variables.  \FIG{fig:jetcom_sgten} and \Fig{fig:jetcom_sgtan} (and also \Fig{fig:jetcom_sgten1app} to \Fig{fig:jetcom_sgtpr2app}) show that in most cases the impact of the
different recombination schemes is small and at the level of a few per
cent only. 
However, there are also pronounced differences: The shapes
of some energy distributions can differ up to $50\%$ in bins near the
kinematical threshold (cf.  \Fig{fig:jetcom_sgten}). This is easily understood by the fact that the
$3\to 2$ prescription strictly keeps the jets on their
mass-shell while the masses of the $2\to 1$ clustered jets
can differ severely from the masses of the parent partons. This is
studied in detail in \Fig{fig:jetcom_sgtm}, where we show the mass
distribution of the jet containing the top quark and the light jet for
the $2\to 1$ clustering. In the $3\to 2$ clustering the
distribution would be proportional to delta functions:
$\delta(J_t^2-m_t^2)$ in case of the top jet and $\delta(J_j^2)$ in
case of the light jet. In particular, at the phase space boundaries
this can lead to significant differences, explaining the large
effects in the threshold region.  In addition, the $3\to 2$
clustering always maintains exact four momentum conservation and in
particular transverse momentum balance also in the case of an
unresolved jet.\footnote{Note that using the P-scheme where the
  recombined momentum is defined as the sum of the 4-momenta of the
  recombined particles 4-momentum conservation is also guaranteed in
  the recombination.  However, particles close to the beam are simply
  dropped and not recombined. In this case the transverse moment is no
  longer balanced.}  The angular distributions do not have a pronounced mass
dependence and as a consequence the corresponding differential
distributions show only minor differences of a few percent in all
bins (cf.  \Fig{fig:jetcom_sgtan}). It is worth mentioning that the aforementioned differences are
formally due to higher orders in perturbation theory.  The large
differences in specific phase space regions may also signal that the
NLO corrections are large in these regions and fixed order
perturbation theory might become unreliable.  Defining the events in terms of variables with
only a weak dependence on the clustering prescription may thus also
help in improving the reliability of the perturbative description of measured observables.\footnote{In all applications usually some reduction on the used variables occurs: Either because some variables are in principle unmeasurable (e. g. neutrino variables) or because the experimental accuracy is not good. The definition of events in terms of four-momenta is then accomplished from the variables by imposing certain kinematics (cf. below).}

\begin{figure}[htbp]
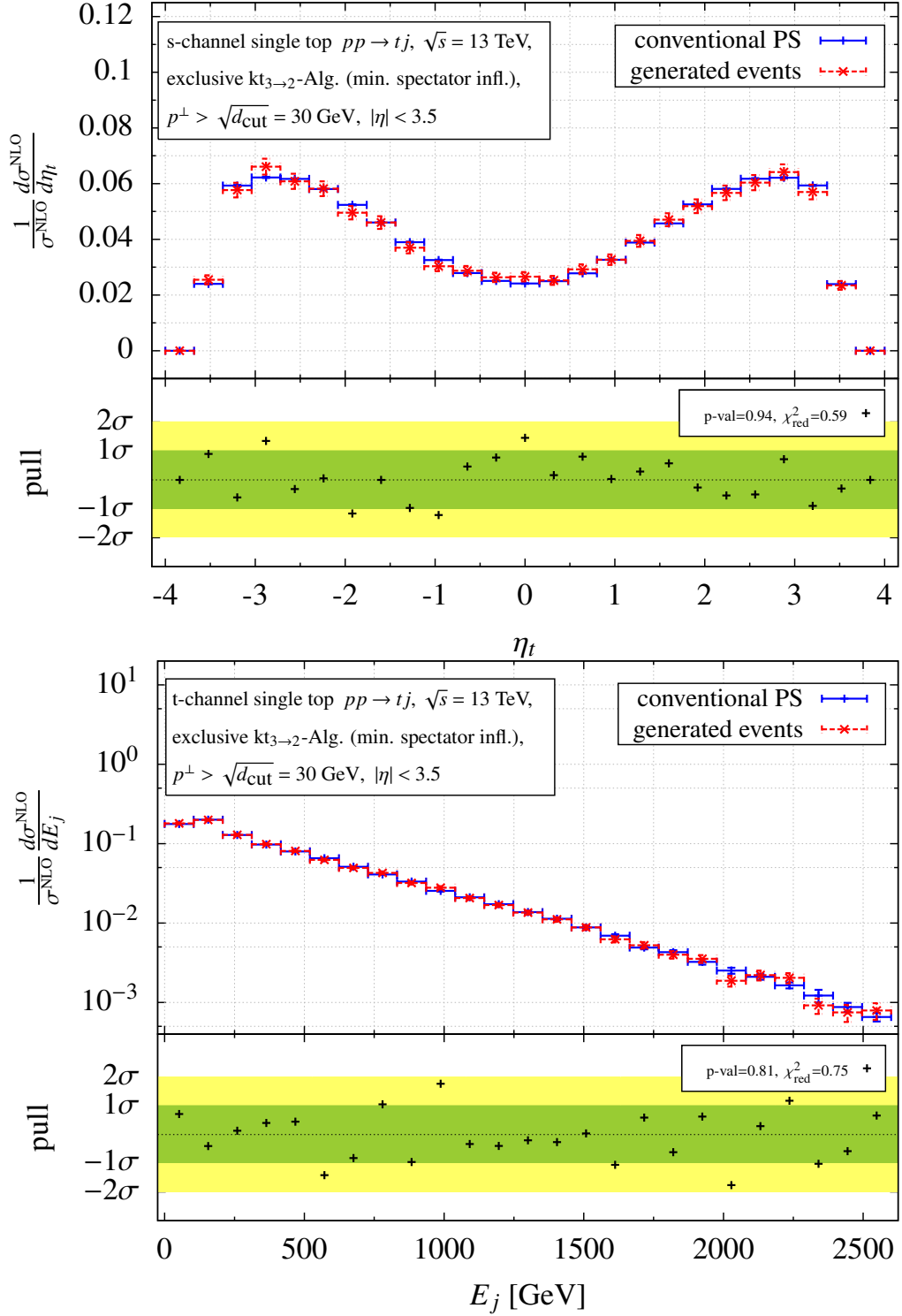

  \begin{flushright}
    \leavevmode
    \includegraphics[height=0.47\textheight]{{{%
          sgtsKT3-2ycut30compmnspecETA1-94evts-crop}}}
    \includegraphics[height=0.47\textheight]{{{%
          sgttKT3-2ycut30compmnspecE2-90evts-crop}}}
    \caption{Pseudo rapidity distribution of the top-tagged jet 
      from exclusive $s$-channel single top production and energy distribution of the light jet 
      from exclusive $t$-channel single top production
      calculated using a conventional parton level MC (blue, solid) compared to 
      histograms filled with generated NLO events (red, dashed).}
    \label{fig:partonMCvsgenev_sgt}
  \end{flushright}
\end{figure}
To define an exclusive single top-quark event in $pp\rightarrow tj$, we
use the pseudo rapidity of the measured top-quark jet ($t$) and  the energy,
pseudo rapidity and azimuthal angle of the light jet ($j$). Since the decay of the top quark is not considered in the calculation
presented here, we use only the rapidity of the top quark and
information related to the light jet. From \Fig{fig:jetcom_sgten} and \Fig{fig:jetcom_sgtan} (and also \Fig{fig:jetcom_sgten1app} to \Fig{fig:jetcom_sgtanapp}) we conclude that for these variables the
impact of the modified clustering prescription is small. The vector
$\vec{x}$ introduced in the previous section, collecting the hadronic
variables, is thus given by
$\vec{x}=(\eta_{t},E_{j}, \eta_{j}, \phi_{j})$. In terms of these variables
the two (resolved) jet momenta read:
\begin{subequations}
\begin{eqnarray}
J_t&=&\left(E_{t},\;-J^{\perp}\cos{\phi_{j}},\;-J^{\perp}\sin{\phi_{j}},
  \;J^{\perp}\sinh{\eta_{t}}\right),\\
J_{j}&=&\left(E_{j},\;J^{\perp}\cos{\phi_{j}},\;J^{\perp}\sin{\phi_{j}},
  \;J^{\perp}\sinh{\eta_{j}}\right)
\end{eqnarray}
\end{subequations}
with
\begin{displaymath}
  J^{\perp}=\frac{E_{j}}{\cosh{\eta_j}}
  \quad\mbox{and}\quad E_{t}=\sqrt{{J^{\perp}}^2\cosh^2{\eta_t}+m^2_t}.
\end{displaymath}
Note that the parameterisation of the jet momenta depends on the top-quark mass which occurs
as a free parameter.
Using
\begin{equation}
  \label{eq:OneParticlePhaseSpaceMeasure}
  d^4p\;\delta(p^2-m^2)
  = \frac{p^{\perp} }{2\cosh{\eta}}\;dE\;d\eta\;d\phi
\end{equation}
and
\begin{eqnarray}
 &&\hspace{-0.5cm} dx_a\;dx_b\;dR_2(J_t,J_j)\nonumber\\
 &=&\frac{1}{2s}\;\frac{{J^{\perp}}^2\cosh{\eta_t}}{E_t\;\cosh{\eta_j}}\;
 dE_j\;d\eta_j\;d\phi_jd\eta_t\;d\vec{J}^{\perp}_t\;dx_a\; dx_b\;
\nonumber \delta\left(\vec{J}^{\perp}_j+\vec{J}^{\perp}_t\right)\\
 &\times& 
 \delta\left(x_a-\frac{1}{\sqrt{s}}(E_j+E_t+J^z_j+J^z_t)\right)\;
 \delta\left(x_b-\frac{1}{\sqrt{s}}(E_j+E_t-J^z_j-J^z_t)\right)
\end{eqnarray}
the event weight
 including NLO corrections reads:
\begin{equation}\label{eq:diffjxsex}
  \frac{d\sigma_\excl^{\text{NLO}}}{d\vec{x}}= 
 \frac{d\sigma_\excl^{\text{NLO}}}{d\eta_{t}\; dE_{j}\; d\eta_{j}\;
   d\phi_{j}}
 =\frac{{J^{\perp}}^2\cosh{\eta_t}}{2\;s\;E_t\;\cosh{\eta_j}} \;
 \frac{d\sigma_\excl^{\text{NLO}}}{dR_2(J_{t},J_{j})}.
\end{equation}
In \Fig{fig:partonMCvsgenev_sgt} the
pseudo rapidity distribution
of the top-tagged and the energy distribution of the light jet are shown for $s$- and $t$-channel
single top-quark production at NLO.  The histograms calculated using
the conventional parton level Monte-Carlo with subsequent
$3\to 2$ jet clustering are shown as blue lines while the
ones filled by the unweighted NLO events are shown as dashed red
lines. Their difference normalised to the statistical uncertainty 
is shown at the bottom of each plot. Again the $p$-value and
the reduced $\chi^2$ of the comparison of the two histograms
are shown. \FIG{fig:partonMCvsgenev_sgt}
illustrates that filling the histograms with the unweighted NLO events
perfectly reproduces the distributions calculated with a conventional
parton level Monte-Carlo integration. Again we have also checked various other distributions which we do not show here. In all cases we find perfect
agreement between the two approaches. We can thus conclude that the
generation of unweighted events with NLO accuracy is successfully
validated.

\subsubsection{Inclusive event definition}
\label{sec:event-definition-inclusive-case}
\begin{figure}[htbp]
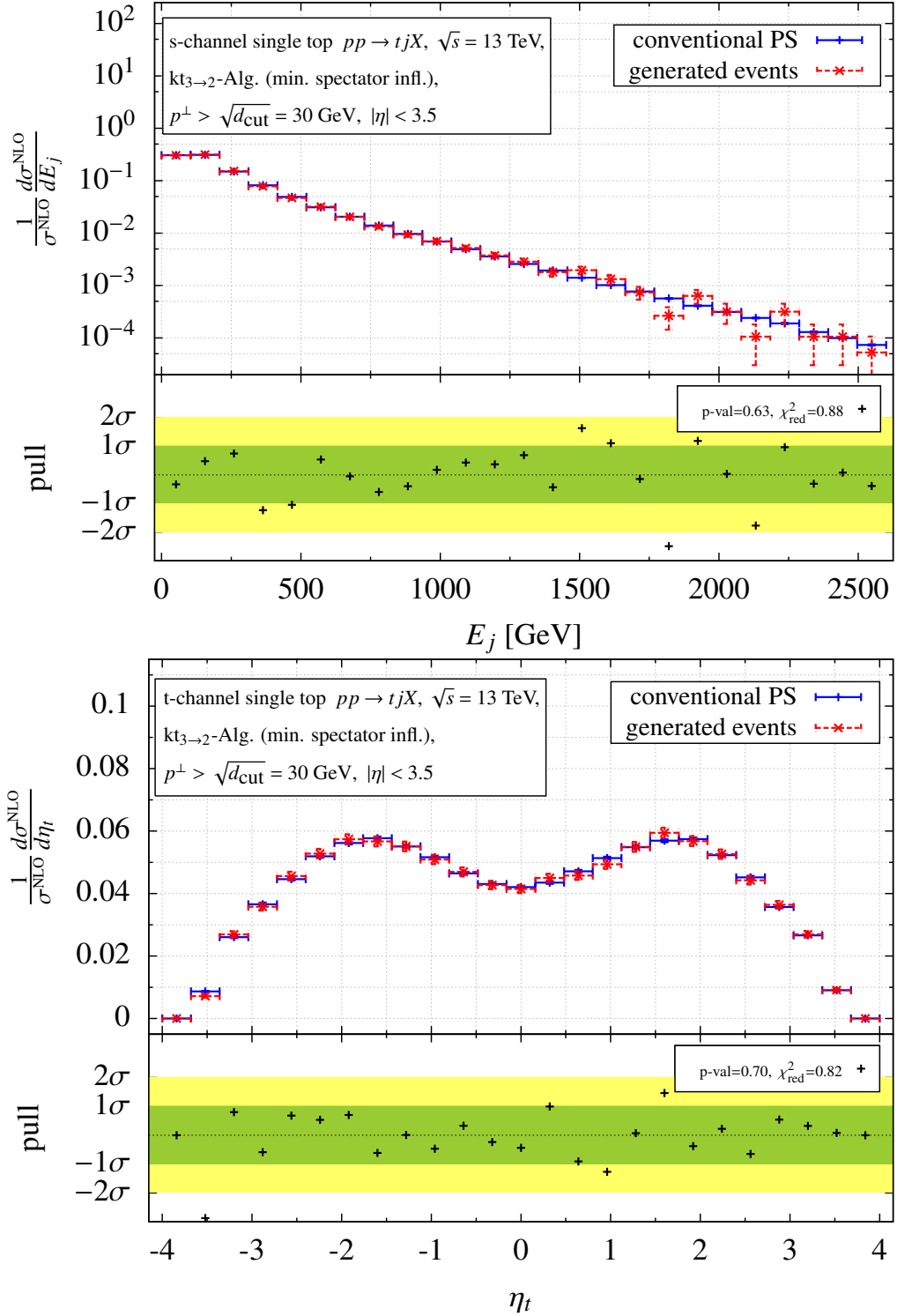

  \begin{flushright}
    \leavevmode
    \includegraphics[height=0.47\textheight]{{{%
          sgtsKT3-2ycut30compmnspecincE2-100evts-crop}}}
    \includegraphics[height=0.47\textheight]{{{%
          sgttKT3-2ycut30compmnspecincETA1-114evts-crop}}}  
    \caption{Energy distribution of the light jet 
      from inclusive $s$-channel single top production and pseudo rapidity distribution of the top-tagged jet 
      from inclusive $t$-channel single top production
      calculated using a conventional parton level MC  (blue, solid) compared to 
      histograms filled with generated NLO events  (red, dashed).}
    \label{fig:partonMCvsgenev_sgtinc}
  \end{flushright}
\end{figure}
In this section we use an inclusive event definition.  We study
$pp\rightarrow tjX$ without a veto on additional jet emission. We note that in the contributions with an additional resolved jet no clustering takes place at NLO accuracy. Taking these contributions into account can therefore only improve the impact of the $3\to 2$ clustering studied in section~\ref{sec:event-definition-exclusive-case} since the
relative weight of clustered events is reduced. Hence, we use
the same set of variables as before: the pseudo rapidity of the
top-quark jet ($t$) and the energy, pseudo rapidity and azimuthal
angle of the hardest light jet ($j$): $\vec{x}=(\eta_{t},E_{j},
\eta_{j}, \phi_{j})$. For events with only one top-tagged jet and one light
jet the same parameterisation as described in the previous section can
be used. In case that an additional jet is resolved (denoted by $J_X$,
$J_j$ denotes the hardest light jet) we use
\begin{subequations}
\begin{eqnarray}
  J_t&=&\left(E_{t},\;-J^{\perp}_t\cos{\phi_{t}},\;
    -J^{\perp}_t\sin{\phi_{t}},\;J^{\perp}_t\sinh{\eta_{t}}\right),\\
  J_{j}&=&\left(E_{j},\;J^{\perp}_j\cos{\phi_{j}},\;J^{\perp}_j
    \sin{\phi_{j}},\;J^{\perp}_j\sinh{\eta_{j}}\right),\\
  J_{X}&=&\left(E_{X},\;J^{\perp}_{X}\cos{\phi_{X}},\;J^{\perp}_{X}
    \sin{\phi_{X}},\;J^{\perp}_{X}\sinh{\eta_{X}}\right)
\end{eqnarray}
\end{subequations}
with
\begin{eqnarray*}
  &J^{\perp}_j=\frac{E_{j}}{\cosh{\eta_j}},\quad J^{\perp}_t
  =\sqrt{(J^{\perp}_j\cos{\phi_{j}}+J^{\perp}_{X}\cos{\phi_{X}})^2
    +(J^{\perp}_j\sin{\phi_{j}}+J^{\perp}_{X}\sin{\phi_{X}})^2},&\\
  &\tan\phi_t=\frac{J^{\perp}_j\sin{\phi_{j}}+J^{\perp}_{X}
    \sin{\phi_{X}}}{J^{\perp}_j\cos{\phi_{j}}+J^{\perp}_{X}\cos{\phi_{X}}},
  \quad
  E_{t}=\sqrt{{J^{\perp}_t}^2\cosh^2{\eta_t}+m^2_t},\quad
  E_{X}=J^{\perp}_{X}\cosh{\eta_X}.&
\end{eqnarray*}
Since we are inclusive in the additional jet the related
variables have to be  integrated over the allowed range:
\begin{displaymath}
  p^{\perp}_{\text{min}}<J^{\perp}_{X}<J^{\perp}_{j},\quad
  0<\phi_{X}<2\pi,\quad 
  -\eta_{\text{max}}<\eta_{X}<\eta_{\text{max}}.
\end{displaymath}
The boundaries follow from the definition of the inclusive event sample.
For the inclusive event definition we choose the central renormalisation
scale $\mu_R$ and factorisation scale $\mu_F$ on an event-by-event
basis as the total transverse energy $E^{\perp}=E\sin\theta
= E^{\perp}_t+E^{\perp}_j+E^{\perp}_X $ of the resolved
final state:
\begin{equation}
  \mu_R=\mu_F=\mu_0=E^{\perp}_t+E^{\perp}_j+E^{\perp}_X.
\end{equation} 
Using again \Eq{eq:OneParticlePhaseSpaceMeasure} and
\begin{eqnarray}
 &&\hspace{-0.5cm} dx_a\;dx_b\;dR_3(J_t,J_j,J_X)\nonumber\\
  \nonumber &=& \frac{1}{4}\; \frac{1}{s}\;
  \frac{{J^{\perp}_j}{J^{\perp}_X}{J^{\perp}_t}\cosh{\eta_t}}
  {E_t\;\cosh{\eta_j}}\;
  dE_j\;d\eta_j\;d\phi_j\;dJ^{\perp}_X\;d\eta_X\;d\phi_X\;d\eta_t\;
  d\vec{J}^{\perp}_t\;dx_a\;dx_b\;
  \\
  \nonumber&\times&\;
  \delta\left(\vec{J}^{\perp}_j+\vec{J}^{\perp}_t+\vec{J}^{\perp}_X\right)\;  
  \delta\left(x_a-\frac{1}{\sqrt{s}}(E_j+E_t+E_X+J^z_j+J^z_t+J^z_X)\right)\\
  &\times&
  \delta\left(x_b-\frac{1}{\sqrt{s}}(E_j+E_t+E_X-J^z_j-J^z_t-J^z_X)\right)
\end{eqnarray}
the $3$-jet contribution to the NLO jet event weight can
be obtained from
\begin{eqnarray}
  \nonumber \frac{d\sigma_{3\text{-jet}}}{d\vec{x}}&=& 
  \frac{d\sigma_{3\text{-jet}}}{d\eta_{t}\; dE_{j}\; d\eta_{j}\; d\phi_{j}}\\
  &=&
  \int\limits_{J^{\perp}_{\text{min}}}^{J^{\perp}_{j}}dJ^{\perp}_{X}
  \int\limits_{-\eta_{\text{max}}}^{\eta_{\text{max}}}d\eta_{X}
  \int\limits_{0}^{2\pi}d\phi_{X}\;
  \frac{{J^{\perp}_j}{J^{\perp}_X}{J^{\perp}_t}
    \cosh{\eta_t}}{4s\;E_t\;\cosh{\eta_j}} \;
  \frac{d\sigma_{3\text{-jet}}}{ dR_3(J_t,J_j,J_X)}.
\end{eqnarray}
The NLO jet event weight for the inclusive event sample is the sum of
the exclusive $2$-jet contribution from \Eq{eq:diffjxsex} and the $3$-jet contribution defined above
\begin{equation}\label{eq:diffjxsin}
 \frac{d\sigma_\incl^{\text{NLO}}}{d\vec{x}}= 
 \frac{d\sigma^{\text{NLO}}_\excl}{d\vec{x}}
 + \frac{d\sigma_{3\text{-jet}}}{d\vec{x}}.
\end{equation}
Again we validate the NLO event generation by comparing differential
distributions calculated using unweighted events with the
distributions determined through a conventional Monte-Carlo
integration. 
\FIG{fig:partonMCvsgenev_sgtinc} shows results of the two
approaches for $s$- and $t$-channel single top-quark
production. As one can see from \Fig{fig:partonMCvsgenev_sgtinc} the two results are in
perfect agreement with each other. We have also scrutinised the other distributions of the observables in the inclusive event definition which we do not show here. Again, we find perfect
agreement between the two approaches. We thus conclude that the generation of
unweighted $tjX$ events including NLO corrections works.

\section{Application: MEM at NLO for single top-quark production}
\label{sec:MEMappl}
In this section we apply the \MEM including NLO corrections to a
hadron collider process.  As an application, we analyse the potential to
measure the top-quark mass using single top-quark events. To do so, we
interpret the unweighted events generated in the previous section as
pseudo data and use the \MEM to extract the top-quark mass. Note that
in the unweighting, the NLO event weight has been used. The simulated event
sample thus incorporates the NLO corrections. 

\subsection{Impact of NLO corrections}
\begin{figure}[htbp]
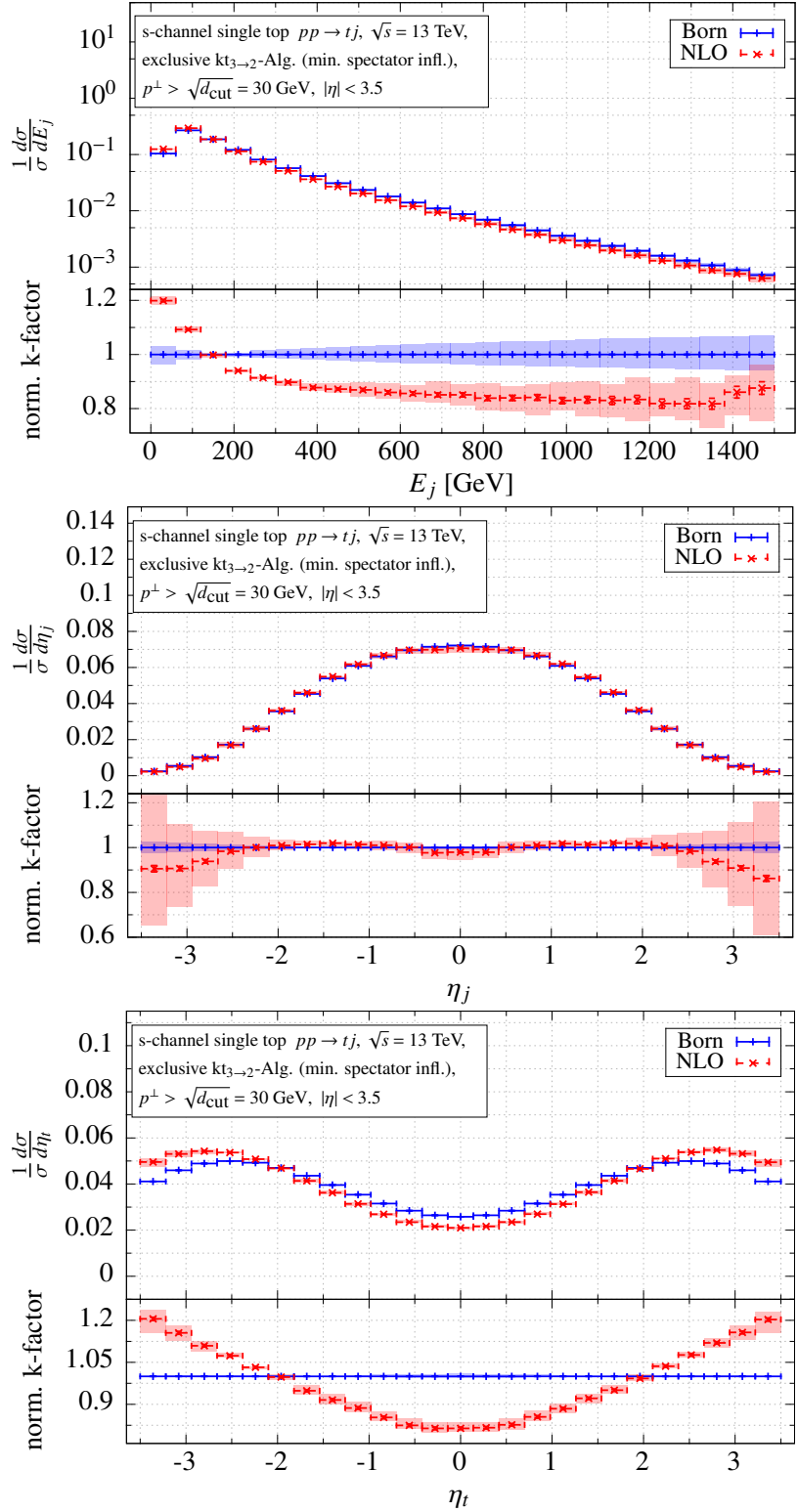

  \begin{center}
    \leavevmode
    \includegraphics[height=0.325\textheight]{{{%
          sgtsKT3-2ycut30compmumnspecE2-200x1e7-crop}}}
    \includegraphics[height=0.325\textheight]{{{%
          sgtsKT3-2ycut30compmumnspecETA2-200x1e7-crop}}}
    \includegraphics[height=0.325\textheight]{{{%
          sgtsKT3-2ycut30compmumnspecETA1-200x1e7-crop}}}
    \caption{Normalised distributions and k-factors for $s$-channel
      single top-quark production using the exclusive event
      definition.}  
    \label{fig:kfac_sgts}
  \end{center}
\end{figure}
\begin{figure}[htbp]
  \begin{center}
    \leavevmode
    \includegraphics[height=0.325\textheight]{{{%
          sgttKT3-2ycut30compmumnspecE2-200x1e7-crop}}}
    \includegraphics[height=0.325\textheight]{{{%
          sgttKT3-2ycut30compmumnspecETA2-200x1e7-crop}}}
    \includegraphics[height=0.325\textheight]{{{%
          sgttKT3-2ycut30compmumnspecETA1-200x1e7-crop}}}
    \caption{Same as \Fig{fig:kfac_sgts} but for the $t$-channel.}  
    \label{fig:kfac_sgtt}
  \end{center}
\end{figure}
\begin{figure}[htbp]
  \begin{center}
    \leavevmode
    \includegraphics[height=0.325\textheight]{{{%
          sgtsKT3-2ycut30compmumnspecincE2-200x1e7-crop}}}
    \includegraphics[height=0.325\textheight]{{{%
          sgtsKT3-2ycut30compmumnspecincETA2-200x1e7-crop}}}
    \includegraphics[height=0.325\textheight]{{{%
          sgtsKT3-2ycut30compmumnspecincETA1-200x1e7-crop}}}
    \caption{Same as \Fig{fig:kfac_sgts}  but for the inclusive event 
      definition. }
    \label{fig:kfac_sgtsinc}
  \end{center}
\end{figure}  
\begin{figure}[htbp]
  \begin{center}
    \leavevmode
    \includegraphics[height=0.325\textheight]{{{%
          sgttKT3-2ycut30compmumnspecincE2-200x1e7-crop}}}
    \includegraphics[height=0.325\textheight]{{{%
          sgttKT3-2ycut30compmumnspecincETA2-200x1e7-crop}}}
    \includegraphics[height=0.325\textheight]{{{%
          sgttKT3-2ycut30compmumnspecincETA1-200x1e7-crop}}}
    \caption{Same as \Fig{fig:kfac_sgtt}  but for the inclusive event 
      definition. }
    \label{fig:kfac_sgttinc}
  \end{center}
\end{figure}  
Before applying the \MEM to simulate a top-quark mass measurement it
is instructive to investigate the size of the NLO corrections.
Tab.~\ref{tab:fidxs} summarises the fiducial Born and NLO cross
sections (with and without additional jet veto) for the $s$- and
$t$-channel production of a single top quark in association with a
light jet.
\begin{table}[htbp]
  \centering
  \caption{Fiducial cross sections.}
  \label{tab:fidxs}
  \def\arraystretch{1.5}
  \begin{tabular}{l|l|l|l|l|}
    \cline{2-5}
    & \multicolumn{2}{c|}{$s$-channel}                         
    & \multicolumn{2}{c|}{$t$-channel}                         \\ \cline{2-5} 
    & \multicolumn{1}{c|}{\excl} 
    & \multicolumn{1}{c|}{\incl} 
    & \multicolumn{1}{c|}{\excl} 
    & \multicolumn{1}{c|}{\incl} \\ \hline
    \multicolumn{1}{|c|}{$\sigma^{\text{Born}}$ [pb]} 
    &\multicolumn{2}{c|}{ $3.093_{-0.099}^{+0.075}$ } 
    & \multicolumn{2}{c|}{ $80.07_{-8.37}^{+6.71}$} \\ \hline
    \multicolumn{1}{|c|}{$\sigma^{\text{NLO}}$ [pb]}
    & $3.057_{-0.017}^{+0.029}$ & $4.071_{+0.087}^{-0.062}$ 
    & $60.03_{-2.34}^{+3.31}$ & $80.45_{-0.33}^{+1.48}$ \\ \hline
\end{tabular}
\end{table}
From Tab.~\ref{tab:fidxs} we see that the fiducial $s$-channel cross section
with a veto on more than one light jet receives a small negative NLO
correction of $-1.2\%$. The contribution of the additional jet adds
$+32.8\%$ to the NLO cross section.  The fiducial $t$-channel cross
section with jet veto on the other hand receives a big negative NLO
correction of $-25.0\%$ while the contribution without jet veto
reduces the size of the NLO correction to $+0.4\%$.

The plots in \Fig{fig:kfac_sgts} to \Fig{fig:kfac_sgttinc} show the impact of the
NLO corrections on the normalised differential distributions. The shaded areas show the simultaneous variation of the renormalisation
and factorisation scale between $\mu_0/2$ and $2\mu_0$ (including
statistical errors).  At the bottom the k-factors which are defined as the ratio
of the NLO results with respect to the Born results are shown. As can be seen
from \Fig{fig:kfac_sgts} and \Fig{fig:kfac_sgtt}, the impact of
the NLO corrections on the shapes of the $s$-channel ($t$-channel)
distributions ranges from $-20\%$ to $+20\%$ ($-10\%$ to $+25\%$).
Varying the renormalisation and factorisation scale results in more
pronounced changes in the NLO than in the Born distributions. This is
especially true in case of the $s$-channel for large values of the
pseudo rapidity. Note that in cases where the real corrections give
large contributions to the NLO corrections---this can happen for
example if a new partonic channel opens---the real corrections may influence
significantly the scale dependence, since this contribution is
essentially leading order and no compensation of the scale variation
takes place (see figures
\ref{fig:kfac_sgtsinc} and \ref{fig:kfac_sgttinc}). 
We conclude that the scale variation of the leading-order results does not offer a reliable estimate for the theoretical uncertainty due to missing higher-order corrections in general.
As can be seen from \Fig{fig:kfac_sgtsinc}, not
vetoing the additional resolved jet results in somewhat flatter
k-factors in regions where most of the events are
expected; except for the $t$-channel pseudo rapidity distribution of the
light jet which receives NLO corrections of up to $+30\%$ in the
central region. \\
When using the extended likelihood in the \MEM the unnormalised
distributions are relevant. For completeness they are given in \Fig{fig:kfac_sgtsapp} to \Fig{fig:kfac_sgttapp} in appendix~\ref{sec:kfactors}.

\subsection{Top-quark mass extraction}
\label{sec:topmassextraction}
\begin{figure}[htbp]
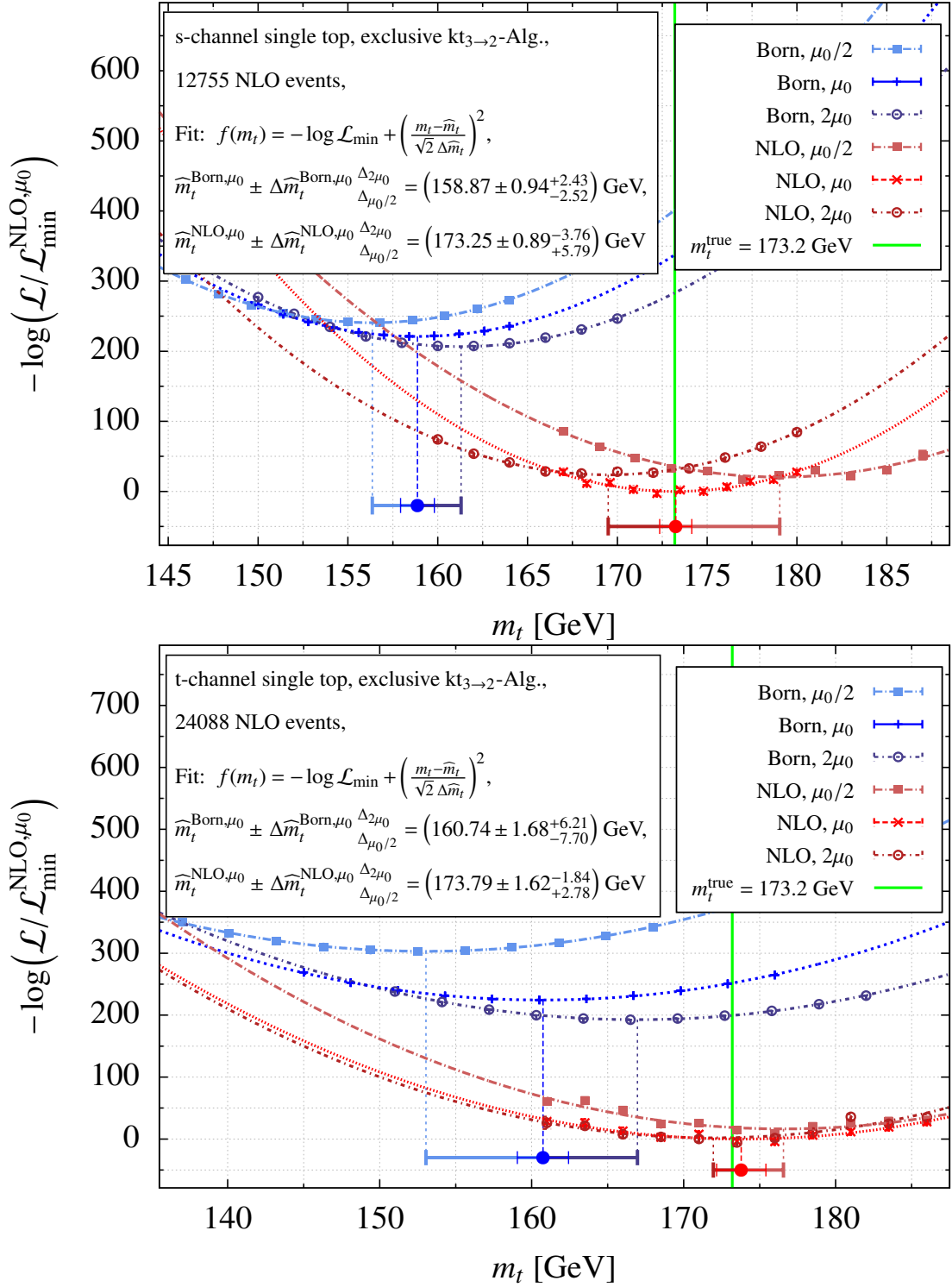

  \begin{center}
    \includegraphics[height=0.47\textheight]{{{%
          sgtsKT3-2ycut30membornnlomu-94evts-crop}}}\\
    \includegraphics[height=0.47\textheight]{{{%
          sgttKT3-2ycut30membornnlomu-90evts-crop}}}
     \caption{Extraction of the top-quark mass from exclusive $s$-channel
       (top) and $t$-channel (bottom) single top-quark events employing NLO (reddish)
       and Born likelihoods (blueish) using a parabola fit around the minimum.}
    \label{fig:MEM_sgts}
  \end{center}
\end{figure}
The NLO jet event weights defined in \Eq{eq:diffjxsex} and \Eq{eq:diffjxsin} can be used to set up likelihood functions as defined in \Eq{eq:likelihood} and extended likelihood functions as defined in \Eq{eq:extLike} at NLO accuracy.
Interpreting the unweighted events generated in section
\ref{sec:event-definition-exclusive-case} and
\ref{sec:event-definition-inclusive-case} as
pseudo data, we now apply the \MEM to them. To investigate the impact of the additional $3$-jet events on the inclusive analysis we choose the relative sizes of the even samples to reflect the fiducial cross sections (cf. table~\ref{tab:fidxs}):
\begin{equation}\label{eq:samsizscal}
N_\incl={\sigma^\NLO_\incl\over\sigma^\NLO_\excl} N_\excl.
\end{equation}
The generated samples of $N_\excl=12755$ ($N_\excl=24088$) NLO exclusive
$s$-channel ($t$-channel) and $N_\incl=16964$ ($N_\incl=32278$) NLO inclusive
$s$-channel ($t$-channel) single top-quark events are thus treated as
results of different (toy) experiments which we analyse using the \MEM
employing the respective (extended) likelihood functions. In particular, as
illustration, the \MEM is used to extract the top-quark mass. This
serves as a consistency check of the procedure and provides another
useful check of the numerical implementation.  Furthermore, the
analysis can be used to estimate the potential of the \MEM applied to
top-quark mass measurements including higher-order corrections.
Figure \ref{fig:MEM_sgts} shows the negative logarithm of the
likelihood (`log-likelihood') as a function of the top-quark mass
evaluated with the sample of $12755$ ($24088$) exclusive NLO
$s$-channel ($t$-channel) single top-quark events.  For each event
sample we calculate the negative logarithm of the likelihood for different values of the top-quark
mass. The results for three different scale settings (see also
\Eq{eq:scale-definition-excl} for the definition of $\mu_0$.) are
shown as open-circles ($\mu=2\mu_0$), crosses ($\mu=\mu_0$) and solid
squares ($\mu=\mu_0/2$).  The lower (red) points are obtained using
NLO predictions for the event weight in the definition of the likelihood,
while the upper (blue) points use the Born approximation for the event
weights.  The
lines connecting the data points show the result of a parabola fit.
The vertical line at $m_t= 173.2\ \GeV$ corresponds to the true value
used in the event generation.  \FIG{fig:MEM_sgts} shows that
including the NLO corrections leads to significant smaller values for
$-\log(\cal{L})$.  The NLO predictions give thus a better description
of the unweighted events. This is not a surprise as the events were generated using the NLO predictions of the cross sections. The good agreement should be seen as a consistency check of the approach.  The top-quark mass is extracted as the
minimum of the negative log-likelihood function. For the
$s$-channel ($t$-channel) we obtain for
$\mu=\mu_0$ an estimator $\widehat{m}^\NLO_t=173.25\ \GeV$ ($\widehat{m}^\NLO_t=173.79\ \GeV$) employing the NLO
likelihood and $\widehat{m}^\Born_t=158.87\ \GeV$ ($\widehat{m}^\Born_t=160.74\ \GeV$) in case the LO likelihood is
used. In addition, we estimate the uncertainty of the determined mass
value. Repeating the mass determination using $\mu=2\mu_0$
and $\mu=\mu_0/2$ a measure for the theoretical uncertainty is calculated from the difference
with respect to the result for $\mu=\mu_0$. The shifts for
$\mu=2\mu_0$ ($\mu=\mu_0/2$) are indicated in the legend as
superscripts $\Delta2\mu_0$ (subscripts $\Delta\mu_0/2$). The scale uncertainty represents a limiting
factor on the reachable accuracy. In addition, we quote the statistical uncertainty of the estimator $\Delta\widehat{m}_t$
determined from the fit.
The results are illustrated as
data points in the lower part of the plots with the uncertainties due to the scale variation shown as thick error bars and the statistical uncertainties shown as thin error bars. While for the $t$-channel
the mass values extracted using LO, respectively NLO predictions are
marginally consistent (taking into account the large scale uncertainty), we find a significant difference of about
$-8.3\%$ in case of the $s$-channel, which is not covered by
the scale variation. 
\begin{figure}[htbp]
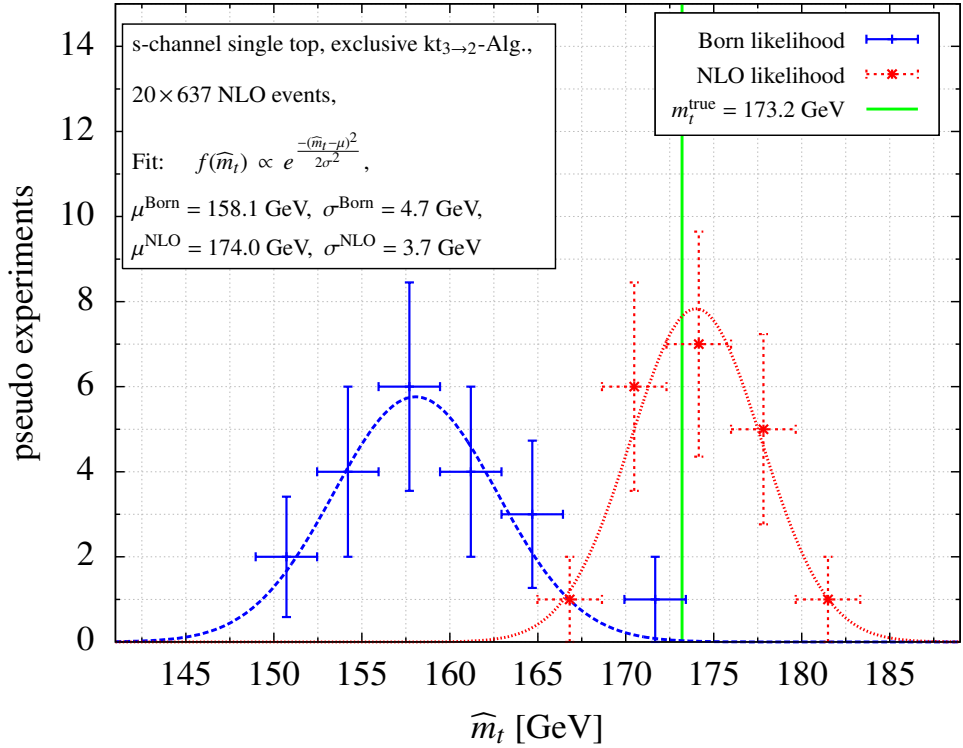

  \begin{center}
    \includegraphics[height=0.47\textheight]{{{%
          sgtsKT3-2ycut30mem-20x637-crop}}}
     \caption{Distribution of the extracted top-quark mass from exclusive
       $s$-channel single top-quark events
       employing NLO (red, dashed) and Born (blue,solid) likelihoods.}
    \label{fig:MEM_sgtsdistr}
  \end{center}
\end{figure}

As mentioned before the Maximum Likelihood
estimator is prone to introduce a bias in case the data is not well
described by the assumed probability distribution. To make sure that
this indeed the origin of the aforementioned discrepancy, we
investigate the consistency of the procedure by splitting the data
sample into $20$ subsamples, treating each subsample as an individual
experiment. Binning the results based on the $20$ subsamples allows to
check whether the bias is of systematic origin as suggested before or
due to some statistical outliers. The result for  the $s$-channel with $637$ events per
subsample is shown in
\Fig{fig:MEM_sgtsdistr}.
As expected, including the NLO corrections in the evaluation of the event weights,
\Fig{fig:MEM_sgtsdistr} shows that
no bias is introduced in the estimator. In contrast, employing only
the Born approximation in the calculation of the event weights, a
significant bias is introduced. \FIG{fig:MEM_sgtsdistr} illustrates
that the shift observed using LO predictions is indeed a consequence of assuming the
`wrong' probability distribution for the event sample and not due to a statistical fluctuation,
since the results of the 20 pseudo experiments give a consistent
picture.  We stress that this shift per se does not preclude the
application of the \MEM in LO: the shift can be accounted for by an
additional calibration procedure. However, since this calibration
introduces additional uncertainties it is preferable make the assumed probability distribution as accurate as possible. While the occurrence of the mass shift using LO or NLO predictions is not surprising we stress that the size of the effect was never precisely quantified before. Furthermore, as mentioned above
estimating the leading-order theory uncertainties using scale
variation does not provide a reliable estimate of higher order effects. 
While the calibration may reproduce  the true mass value, the
uncertainty estimates would be unreliable. In particular, compared to
the NLO result the uncertainties could be underestimated.

We shall also comment on the scale uncertainty observed using LO and
NLO predictions. While in the $t$-channel the uncertainty due to the
scale variation is reduced when going from LO to NLO predictions, this
is not the case for the $s$-channel. Varying the scale in the
likelihood used to extract the top-quark mass results in a shift of
the NLO estimator between $-2.2\%$ and $+3.3\%$ while the Born
estimator varies between $-1.6\%$ and $+1.5\%$.  For the $s$-channel,
the related uncertainties thus increase when going from LO to NLO, in
contrast to the naive expectation. However, the LO predictions are due
to electroweak interaction and the scale enters only through the
factorisation scale dependence of the parton distribution functions.
Furthermore, since the event weight is calculated from normalised
distribution, this dependence cancels to some extent in the ratio.
As a consequence, the scale variation of the leading-order result does not provide a reliable method to estimate the effect of
higher-order corrections.  It is thus not surprising that the picture
we observe is different from the naive expectation and the
uncertainties based on the LO scale variation underestimates the true uncertainty. We also point out that for $s$-channel
and $t$-channel production the scale uncertainties dominate.  The
statistical uncertainties amount to $0.9$ \GeV for $s$-channel and $1.6$
\GeV for the $t$-channel (in the exclusive $s$-channel ($t$-channel)
analysis the event sample contains only $12755$ ($24088$) events). 
\paragraph{Inclusive event definition}
\begin{figure}[htbp]
  \begin{center}
    \includegraphics[height=0.47\textheight]{{{%
          sgtsKT3-2ycut30membornnloincmu-100evts-crop}}}\\
    \includegraphics[width=0.99\textwidth]{{{%
          sgttKT3-2ycut30membornnloincmu-114evts-crop}}}
     \caption{Same as \Fig{fig:MEM_sgts} but for the inclusive event definition.}
    \label{fig:MEM_sgtsinc}
  \end{center}
\end{figure}
Note that vetoing additional resolved jets requires phase space cuts on the additional radiation. These cuts introduce additional scales into the NLO calculation leading to potentially uncancelled logarithms which might spoil the convergence of perturbation theory. In the inclusive event definition this veto is dropped.
\FIG{fig:MEM_sgtsinc} shows the corresponding analysis using
the inclusive event definition.  In difference to the exclusive case
discussed before, we observe for both production channels an
improvement of the scale uncertainties when going from LO to NLO. For
the $t$-channel, the uncertainties are almost reduced by a factor of
ten.  In case of the $s$-channel the improvement is less impressive.
The scale variation of $-1.7\%$ and $+1.4\%$ for the Born estimator is
reduced to $-0.9\%$ and $+1.3\%$ in case the NLO likelihood is used.
We also note that for the inclusive event definition the difference of
$-1.3\%$ between the central values in LO and NLO is now well covered by the
scale variation. For
both production channels, we find that the extracted top-quark mass is
consistent with the true value used in the event generation.  While in
the $s$-channel the scale uncertainty dominates, the scale uncertainty
in the $t$-channel amounts to less than $1\ \GeV$ uncertainty of the
extracted top-quark mass and the total uncertainty is dominated by the
statistical uncertainty. The latter could be reduced using a larger
event sample. We also note that the increased sample size of the inclusive events with respect to the exclusive event sample (cf. \Eq{eq:samsizscal}) does not result in improvements of the relative uncertainties of the analyses. As presented further down in \Eq{eq:fidxsmassexcl} and \Eq{eq:fidxsmassincl}, the inclusive cross section is slightly less sensitive to the top-quark mass which compensates the statistical gain of the larger sample sizes.
\paragraph{Extended likelihood}
\begin{figure}[htbp]
  \begin{center}
    \includegraphics[height=0.47\textheight]{{{%
          sgtsKT3-2ycut30membornnlomu-94evts_ext-crop}}}\\
    \includegraphics[height=0.47\textheight]{{{%
          sgttKT3-2ycut30membornnlomu-90evts_ext-crop}}}
     \caption{Same as  \Fig{fig:MEM_sgtsdistr} but using the extended
       likelihood.
       }
     \label{fig:MEM_sgts_ext}
  \end{center}
\end{figure}
\begin{figure}[htbp]
  \begin{center}
    \leavevmode
    \includegraphics[height=0.47\textheight]{{{%
          sgtsKT3-2ycut30membornnloincmu-100evts_ext-crop}}}\\
    \includegraphics[height=0.47\textheight]{{{%
          sgttKT3-2ycut30membornnloincmu-114evts_ext-crop}}}
     \caption{Same as \Fig{fig:MEM_sgtsinc} but using the extended likelihood
       .}
    \label{fig:MEM_sgtsinc_ext}
  \end{center}
\end{figure}
In the previous section we argued that with extended likelihoods the information contained in the total event number may be used to improve the parameter determination.
\FIG{fig:MEM_sgts_ext} (\FIG{fig:MEM_sgtsinc_ext}) shows the
extended log-likelihoods for $s$-channel and $t$-channel production
using the exclusive (inclusive) event definition. As before we observe
that using the NLO likelihood correctly reproduces the true mass value
used in the event generation. 
Using the extended Born likelihood the values of the estimators are driven by two competing effects: Reproducing both the NLO distributions and the NLO fiducial cross sections using only Born predictions. When comparing the exclusive analyses presented in \Fig{fig:MEM_sgts} and \Fig{fig:MEM_sgtsinc} with the fiducial cross sections given in table~\ref{tab:fidxs}, the analyses based on extended likelihoods show that the impact of the expected total event numbers is dominating this compromise: In the exclusive case the fiducial $s$-channel cross section receives only a small NLO correction which leaves little room for a shift of the top-quark mass. On the other hand, the exclusive fiducial $t$-channel cross section receives a large negative NLO correction. Hence, the estimator extracted with the extended Born likelihood gets pushed to higher top-quark mass values in order to reduce the Born cross section accordingly. In the inclusive analysis the $3$-jet contribution significantly increases the NLO correction to the fiducial $s$-channel cross section. As a result, the Born estimator is pushed to smaller top-quark mass values to open up more phase space for the Born cross section. In the inclusive $t$-channel the $3$-jet contribution compensates the large negative exclusive NLO correction leaving only little room for a shift in the Born estimator. The shifts of the NLO and Born estimators due to scale variation in the extended likelihoods largely follow the same pattern deduced from table~\ref{tab:fidxs} which overall leads to a considerable reduction of the theoretical uncertainty estimates when taking NLO corrections into account.
Concentrating on the statistical
uncertainties, we observe a significant improvement in case the extended likelihood is used: The uncertainty is
roughly reduced by a factor of two.
\begin{figure}[htbp]
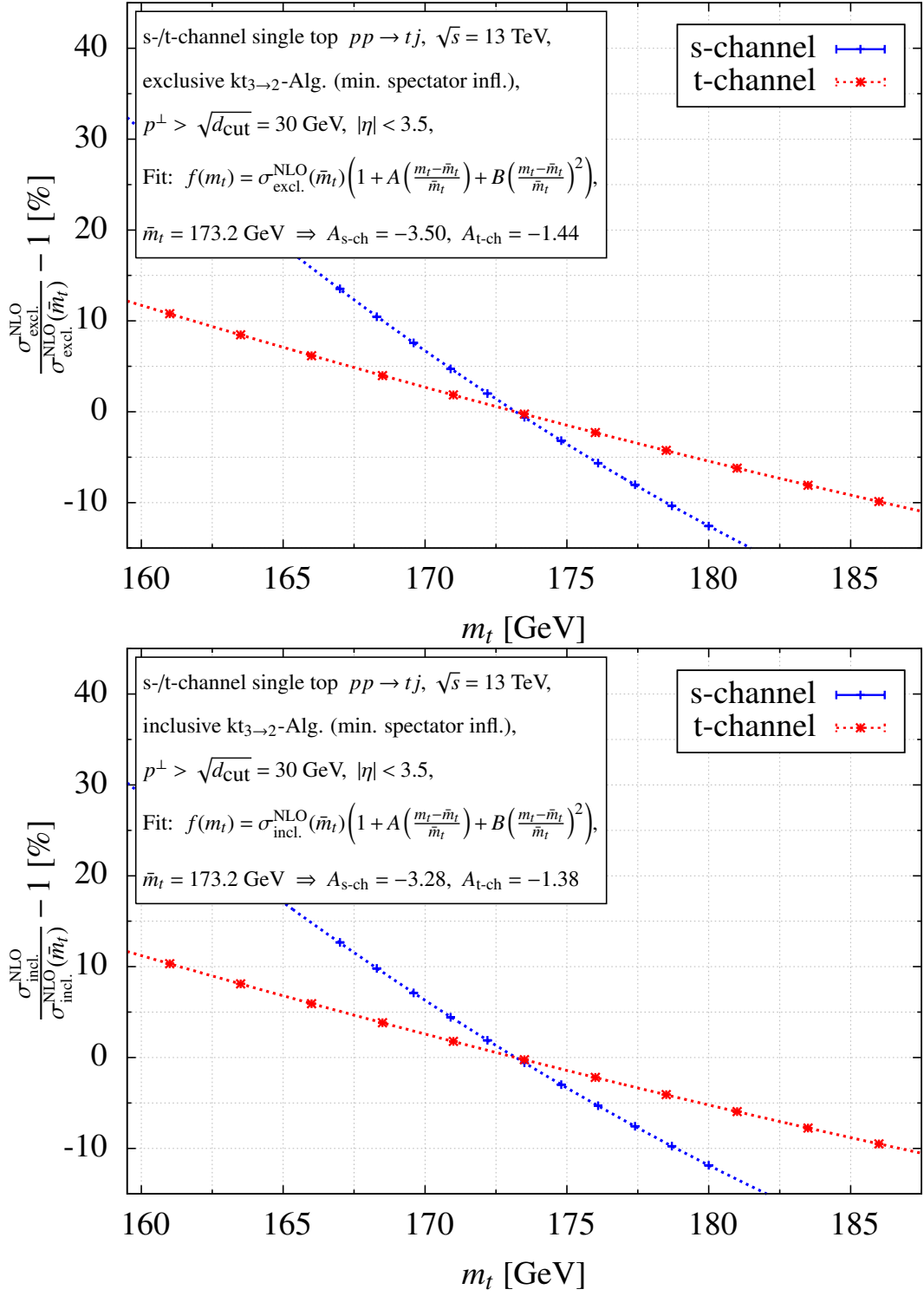

  \begin{center}
    \includegraphics[height=0.47\textheight]{{{%
          sgttKT3-2ycut30sgt-excl-crop}}}
    \includegraphics[height=0.47\textheight]{{{%
          sgttKT3-2ycut30sgt-incl-crop}}}
     \caption{Top-quark mass dependence of the fiducial exclusive (top) and inclusive (bottom) $s$- and $t$-channel cross sections calculated at NLO accuracy.}
    \label{fig:MEM_sgtmass}
  \end{center}
\end{figure}
The relative uncertainty fits
nicely with the naive expectation based on the assumption that the
cross section will be measured with an relative uncertainty of the
order $1/\sqrt{N}$ where $N$ denotes the total event number:
In \Fig{fig:MEM_sgtmass} we show the dependence of the fiducial exclusive and inclusive cross sections calculated including NLO corrections and infer the top-quark mass sensitivities by approximating the dependence of the cross section on the top-quark mass by a polynomial as presented in \Refs{Kant:2014oha,Campbell:2009gj}. For the top-quark mass sensitivities of the fiducial exclusive cross sections this yields
\begin{equation}\label{eq:fidxsmassexcl}
  \frac{\Delta\sigma^\NLO_{\excl\;s\text{-ch}}}{\sigma^\NLO_{\excl\;s\text{-ch}}} = -3.50\times
  \frac{\Delta\mt}{\mt}\,
  ,\quad
  \frac{\Delta\sigma^\NLO_{\excl\;t\text{-ch}}}{\sigma^\NLO_{\excl\;t\text{-ch}}} = -1.44\times
  \frac{\Delta\mt}{\mt}\,
\end{equation} 
The fiducial inclusive cross sections are slightly less sensitive to the mass of the top-quark
\begin{equation}\label{eq:fidxsmassincl}
  \frac{\Delta\sigma^\NLO_{\incl\;s\text{-ch}}}{\sigma^\NLO_{\incl\;s\text{-ch}}} = -3.28\times
  \frac{\Delta\mt}{\mt}\,
  ,\quad
  \frac{\Delta\sigma^\NLO_{\incl\;t\text{-ch}}}{\sigma^\NLO_{\incl\;t\text{-ch}}} = -1.38\times
  \frac{\Delta\mt}{\mt}\,. 
\end{equation}
From \Eq{eq:fidxsmassexcl} and \Eq{eq:fidxsmassincl} we can estimate the relative uncertainties of top-quark mass determinations based on idealised measurements of fiducial cross sections as
\begin{equation}\label{eq:hypmassmeas}
\left|{\Delta\mt^{\excl\;s\text{-ch}}\over m_t}\right|=\underbrace{{\Delta\sigma^\NLO_{\excl\;s\text{-ch}}\over\sigma^\NLO_{\excl\;s\text{-ch}}}}_{={12755}^{-1/2}}\,{1\over 3.50} = 0.25\%
  ,\quad
\left|{\Delta\mt^{\excl\;t\text{-ch}}\over m_t}\right|=\underbrace{{\Delta\sigma^\NLO_{\excl\;t\text{-ch}}\over\sigma^\NLO_{\excl\;t\text{-ch}}}}_{={24088}^{-1/2}}\,{1\over 1.44} = 0.45\%
\end{equation}
and
\begin{equation}\label{eq:hypmassmeasincl}
\left|{\Delta\mt^{\incl\;s\text{-ch}}\over m_t}\right|=\underbrace{{\Delta\sigma^\NLO_{\incl\;s\text{-ch}}\over\sigma^\NLO_{\incl\;s\text{-ch}}}}_{={16964}^{-1/2}}\,{1\over 3.28} = 0.23\%
  ,\quad
\left|{\Delta\mt^{\incl\;t\text{-ch}}\over m_t}\right|=\underbrace{{\Delta\sigma^\NLO_{\incl\;t\text{-ch}}\over\sigma^\NLO_{\incl\;t\text{-ch}}}}_{={32278}^{-1/2}}\,{1\over 1.38} = 0.40\%.
\end{equation}
To estimate the impact of the information contained in the event sample sizes ($N_\excl$ and $N_\incl$) we form weighted averages of the results given in \Eq{eq:hypmassmeas} and \Eq{eq:hypmassmeasincl} and the results of the analyses presented in \Fig{fig:MEM_sgts} and \Fig{fig:MEM_sgtsinc} (which are only sensitive to normalised differential cross sections).  We find relative uncertainties for the respective weighted averages of $\Delta\overline{m}_t/\overline{m}_t=\pm 0.23\%$ ($\pm 0.40\%$) in case of the exclusive $s$-channel ($t$-channel) and $\Delta\overline{m}_t/\overline{m}_t=\pm 0.21\%$ ($\pm 0.37\%$) in case of the inclusive $s$-channel ($t$-channel). These improved top-quark mass sensitivities achieved by this simple combination match the relative uncertainties of the extended likelihood analysis presented in \Fig{fig:MEM_sgts_ext} and \Fig{fig:MEM_sgtsinc_ext}.
 As mentioned before this gain in
sensitivity should be compared with the additional systematic uncertainty
due to the imperfect knowledge of the integrated luminosity. To
estimate the impact of the uncertainty of the luminosity, we have repeated the
analysis varying the luminosity by $\pm \Delta L/{L}\approx \pm2\%$
(see e.g. \Ref{Aaboud:2016hhf}) . We identify the observed
shifts of the extracted top-quark masses of about $\pm 1$ GeV ($\pm 2$
GeV) for the $s$-channel ($t$-channel) as an additional systematic
uncertainty. Because of this additional uncertainty the extended
likelihood does not lead to a more precise measurement unless the
uncertainty of the integrated luminosity is significantly reduced.

The results from the various analyses shown in \Fig{fig:MEM_sgts} to
\Fig{fig:MEM_sgtsinc_ext} are summarised in table~\ref{tab:schmasses} and table~\ref{tab:tchmasses}.  As mentioned before, using the NLO likelihood
always reproduces the true mass value $m^{\text{true}}_t=173.2\ \GeV$
used in the event generation.  Using instead the LO likelihood to
extract the top-quark mass can lead to significant deviations from the
true value. These deviations tend to be smaller for the inclusive
event definition. In most cases the inclusion of the NLO corrections
reduces the uncertainty related to the scale variation. However, as
mentioned before, one should keep in mind that the LO prediction is a
purely electroweak process and the variation of the factorisation
scale alone does not give a reliable estimate of higher-order corrections.
This is also reflected in the observation that for the $s$-channel the
mass values extracted from the exclusive event sample using LO and NLO
predictions do not agree within the scale uncertainty.
\begin{table}[htbp]
\centering
\caption{Extracted top-quark mass estimators from NLO $s$-channel events.}
\label{tab:schmasses}
\def\arraystretch{1.8}
\begin{tabular}{|c|c|c|c|}
\cline{3-4}
\multicolumn{2}{c|}{}&\multicolumn{2}{c|}{estimator $\pm\text{ stat. err. }^{\text{scale var.} \uparrow}_{\text{scale var.} \downarrow}$ [GeV]}\\ \cline{3-4}
\multicolumn{2}{c|}{}& {$N_\excl=12755$} & {$N_\incl=16964$} \\ \hline
\multirow{2}{*}{Likelihood} & {NLO} & {$173.25\pm0.89^{-3.76}_{+5.79}$} & {$173.07\pm0.90^{-1.61}_{+2.32}$} \\ \cline{2-4}
&{Born} & {$158.87\pm0.94^{+2.43}_{-2.52}$} & {$171.01\pm0.92^{+2.42}_{-2.89}$} \\ \hline
\multirow{2}{*}{Extended Likelihood} &{NLO } &{$173.35\pm0.40^{-0.23}_{+0.65}$} & {$173.22\pm0.37^{-0.80}_{+1.39}$} \\ \cline{2-4}
&{Born} & {$171.22\pm0.41^{+1.62}_{-1.97}$} & {$160.88\pm0.35^{+1.53}_{-2.06}$} \\ \hline
\end{tabular}
\end{table}

\begin{table}[htbp]
\centering
\caption{Extracted top-quark mass estimators from NLO $t$-channel events.}
\label{tab:tchmasses}
\def\arraystretch{1.8}
\begin{tabular}{|c|c|l|l|}
\cline{3-4}
\multicolumn{2}{c|}{}&\multicolumn{2}{c|}{estimator $\pm\text{ stat. err. }^{\text{scale var.} \uparrow}_{\text{scale var.} \downarrow}$ [GeV]}\\ \cline{3-4}
\multicolumn{2}{c|}{}& {$N_\excl=24088$} & {$N_\incl=32278$} \\ \hline
\multirow{2}{*}{Likelihood} & {NLO} &{$173.79\pm1.62^{-1.84}_{+2.78}$} & {$173.85\pm1.62^{+0.84}_{-0.46}$} \\ \cline{2-4}
&{Born} &{$160.74\pm1.68^{+6.21}_{-7.70}$} & {$168.88\pm1.47^{+6.26}_{-7.76}$} \\ \hline
\multirow{2}{*}{Extended Likelihood} &{NLO} & {$173.53\pm0.70^{+4.70}_{-3.39}$} & {$173.53\pm0.65^{+2.03}_{-0.51}$} \\ \cline{2-4}
&{Born} & {$201.43\pm0.73^{+8.89}_{-12.50}$} & {$172.37\pm0.58^{+8.65}_{-12.31}$} \\ \hline
\end{tabular}
\end{table}
\begin{figure}[htbp]
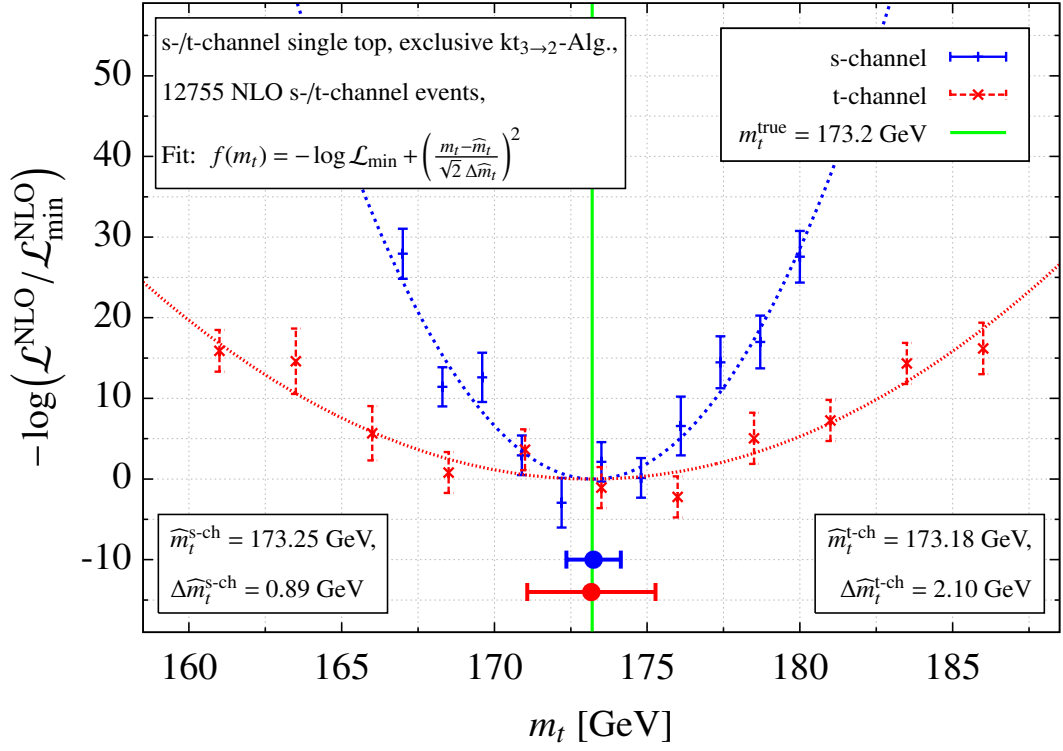

  \begin{center}
    \includegraphics[height=0.47\textheight]{{{%
          sgtKT3-2ycut30memdet_st-crop}}}
     \caption{Extraction of the top-quark mass for a fixed 
       number of NLO $s$- and $t$-channel single top-quark events.}
    \label{fig:MEM_sgt}
  \end{center}
\end{figure}
\begin{figure}[htbp]
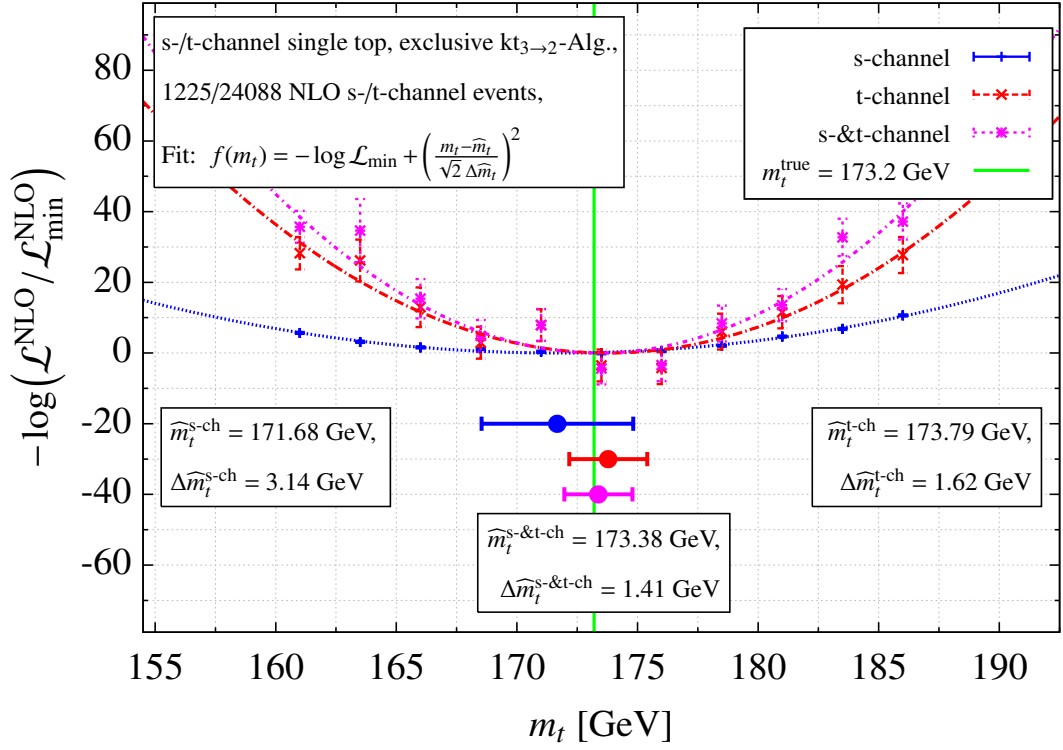

  \begin{center}
    \includegraphics[height=0.47\textheight]{{{%
          sgtKT3-2ycut30memdet_strel-crop}}}
     \caption{Extraction of the top-quark mass for a fixed 
       luminosity.}
    \label{fig:MEM_sgtrel}
  \end{center}
\end{figure}
In \Fig{fig:MEM_sgt} we compare the sensitivity to the top-quark mass for $s$- and $t$-channel single
top-quark production for a fixed number of events. The $s$-channel shows a
higher mass sensitivity leading as far as the statistical
uncertainties are concerned to a more precise mass measurement assuming the same number of events. Because
of the higher mass sensitivity of the $s$-channel, the log-likelihood
is significantly narrower than the log-likelihood for the $t$-channel
leading to a reduction of the statistical uncertainty by roughly a
factor $2.3$ compared to the $t$-channel. This factor is also consistent with
the results presented for the fiducial cross sections in \Eq{eq:fidxsmassexcl}.
In practice, the higher sensitivity of the $s$-channel is compensated
by the $20$ times larger event rate of the $t$-channel. It is thus
interesting to study the impact of the $s$-channel on a mass
measurement assuming a realistic mixture of $s$- and $t$-channel events. 
\FIG{fig:MEM_sgtrel} illustrates the mass measurement
using an event sample with $24088$ ($1225$) $t$-channel ($s$-channel)
events. As one can see in \Fig{fig:MEM_sgtrel} the $t$-channel leads to a
significantly smaller uncertainty in the mass extraction. The
uncertainty in the $s$-channel is roughly twice as large as in the
$t$-channel.
Including both channels in the mass extraction leads to a minor
improvement of the statistical uncertainty by about $200$ MeV compared to the $t$-channel only case.
This observation is again consistent with a simple weighted mean of the separate mass extractions from $24088$ $t$-channel events and $1225$ $s$-channel events yielding $\overline{m}_t\pm\Delta\overline{m}_t=173.35\pm1.44$ GeV.

\section{Conclusion}\label{sec:concl}
In this article we have presented the first application of the MEM at
NLO to the hadronic production of jets using the example of single
top-quark production at the LHC.

Modifying the recombination prescription of conventional $2\to 1$ jet
algorithms by using a $3\to 2$ clustering inspired by the
Catani-Seymour dipole formalism allows to factorise the real phase
space such that the unresolved phase space regions can be efficiently
integrated numerically.  Because of the $3\to 2$ clustering the
clustered jet momenta satisfy four-momentum conservation and the
on-shell conditions. The momenta of the resolved jets can thus be used
in the virtual matrix element enabling a point-wise cancellation of IR
divergences in combination with the real contribution---assuming an IR
regularisation scheme has been used in intermediate steps. It is thus
possible to define a fully differential event weight for jets instead
of partons.  As an important application the event weight can be used
to generate unweighted jet events distributed according to the NLO
cross section. It is also possible to attribute NLO weights to jet
events observed in experiments---as needed for the MEM at NLO.

As proof of concept, we have generated events following the NLO cross
section for the hadronic $s$- and $t$-channel production of a single
top quark in association with a light jet (with an optional veto on
additional jet activity). We have checked that the unweighted events
reproduce the differential distributions obtained using a conventional
parton level Monte-Carlo integration.  As a further consistency check,
the top-quark mass has been successfully reproduced from the event
samples using the MEM at NLO. When extracting the top-quark mass using
Born matrix elements in the MEM we find significant biases in the
estimators. In particular, we observe that in general the shift is not
covered by the theoretical uncertainties estimated using scale
variation. While the shift could be taken into account through a
calibration procedure most likely the theoretical uncertainties would
be underestimated relying on LO results only.  Using the \MEM at NLO
accuracy might thus help in reducing the required calibration
(together with the associated uncertainty) and providing a reliable
estimate of the theoretical uncertainty. Furthermore, when using the
\MEM at NLO accuracy the renormalisation scheme of the extracted
parameters is uniquely defined.

We have applied the general algorithm presented in
\Ref{Martini:2015fsa} which relies on a modification of the clustering
prescription to cover the most general event definitions in terms of
any observables that can be constructed from the $4$-momenta of the
observed jets.

We stress that the analyses presented in this paper only present the
starting point for the MEM at NLO machinery. An important aspect of
this article is to serve as a proof of concept and illustrate the
potential gain of using then \MEM at NLO accuracy.
Further studies are required to describe more realistic final states:
While the inclusion of the top-quark decay in the calculation of the
likelihood should be conceptually straightforward more work is needed
to incorporate also the effect of a parton shower. (The
  possibility to combine the approach pursued in
  \Refs{Soper:2011cr,Soper:2014rya} with the MEM at NLO framework
  presented here could for example be investigated.) For more
complicated processes the \PSS used in this
article to cancel the IR singularities, should be replaced by a subtraction
type method. As already stated in section~\ref{sec:MEM-Basics},
considering non-trivial transfer functions in the framework of the
\MEM at NLO accuracy might require careful studies and dedicated
research on its own. With all that in mind, let us point out that the
performance of the \MEM at NLO and LO accuracy should ultimately be
compared in the application to events that mimic realistic outcomes of
experiments as closely as possible to get the best estimate on what
would happen with real data.  Nevertheless, despite some limitations the 
studies presented in this work illustrate that missing
higher-order corrections in the theoretical predictions might not only
introduce biases in the extracted estimators but may also render the
inferred systematic uncertainty estimates unreliable. Under the
assumption that fixed-order NLO calculations give a better
approximation of real data than LO ones, we expect these effects to
play a role in the experimental application of the \MEM as well.  For
future studies it will be worthwhile to repeat the corresponding
analyses with events obtained from an independent, state-of-the-art
NLO event generator with a parton shower that can be tuned to data
like e.g. POWHEG (see \Ref{Frixione:2007vw}) or MG5$\_$aMC@NLO (see
\Ref{Alwall:2014hca}). This will allow to further quantify whether effects
of similar size to the above findings are also to be expected in 
parameter extractions when the method is applied to real data.

\section*{Acknowledgements}
We would like to thank Markus Schulze for his careful reading of the manuscript and his valuable comments. We acknowledge useful discussions with our colleagues of the ATLAS collaboration, in particular: Patrick Rieck, S{\"o}ren Stamm and Oliver Kind. This work is supported by the German Federal Ministry for Education and Research (grant 05H15KHCAA).

\appendix

\renewcommand*{\thesection}{\Alph{section}}
\section{Phase space parameterisation adapted to the \PSS}
\label{sec:ps-parameterisation}
\subsection{Final-state clustering with final-state spectator}
\label{sec:ffm}
For massive partons $i$, $j$, $k$ the phase space can be factorised
in terms of a phase space of $n$ jets and the dipole phase space 
measure $dR_{ij,k}$ related to the clustered parton:
\begin{equation}
  dR_{n+1}\left(P,p_1,\ldots,p_{n-2},p_i,p_k,p_{j}\right)
  = dR_{n}\left(P,J_1,\ldots,J_{n-2},J_j,J_{k}\right)dR_{ij,k}.
\end{equation}

When employing the \PSS it is useful to express the integration via the
invariants 
\begin{equation}
  Q^2=\left(J_j+J_k\right)^2, \quad 
  s'_{ij}=2p_i\cdot p_j,\quad
  s_{ij}=s'_{ij}+m^2_i+m^2_j, \quad  
  s'_{ik}=2p_i\cdot p_k,
\end{equation}
with
\begin{equation}
 y=\frac{s'_{ij}}{Q^2-m^2_i-m^2_j-m^2_k} 
\end{equation}
and
\begin{equation}
 z=\frac{s'_{ik}}{Q^2-m^2_k-s_{ij}} .
\end{equation}
as
\begin{eqnarray}
  \nonumber dR_{ij,k} &=&
  \frac{1}{32\pi^{3}\sqrt{\Kallen{Q^2,J^2_j,m^2_k}}}d\phi\;ds'_{ij}\;
  ds'_{ik}\Theta\left(\phi\left(2\pi-\phi\right)\right)\\
  &&\times\;\Theta\left(\left(s'_{ij}-{s'_{ij}}^{-}\right)
    \left({s'_{ij}}^{+}-s'_{ij}\right)\right)\Theta
  \left(\left(s'_{ik}-{s'_{ik}}^{-}\right)\left({s'_{ik}}^{+}
      -s'_{ik}\right)\right).
\end{eqnarray}
The K\"all\'en function $\lambda$ is defined as
\begin{equation}
\label{eq:Kallen}
\lambda(x,y,z)=x^2+y^2+z^2 - 2xy - 2xz - 2yz.
\end{equation}
The integration boundaries read
\begin{eqnarray*}
&{s'_{ij}}^{-}=2m_im_j,\quad {s'_{ij}}^{+}=\left(|Q|-m_k\right)^2-m^2_i-m^2_j,&\\
&{s'_{ik}}^{\mp}=(Q^2-m^2_k-s_{ij})\frac{(2m_i^2+s'_{ij})(1\mp v_{ij,i}v_{ij,k})}{2s_{ij}},&
\end{eqnarray*}
with the relative velocities between $p_i+p_j$ and $p_i$ or $p_k$ given by
\begin{eqnarray*}
  &\varv_{ij,i}=\frac{\sqrt{\left(Q^2-m^2_i-m^2_j-m^2_k\right)^2y^2-4m^2_im^2_j}}{\left(Q^2-m^2_i-m^2_j-m^2_k\right)y+2m^2_i},\quad\varv_{ij,k}=\frac{\sqrt{\left[2m^2_k+\left(Q^2-m^2_i-m^2_j-m^2_k\right)\left(1-y\right)\right]^2-4m^2_k}}{\left(Q^2-m^2_i-m^2_j-m^2_k\right)\left(1-y\right)}.&
\end{eqnarray*}
The phase space parameterisation corresponds to the following
clustering of $n+1$ partons to $n$ jets
\begin{eqnarray}\label{eq:ffclus}
\nonumber  J_{k} &=&\left[\sqrt{\frac{\lambda(Q^2,J^2_j,m^2_k)}{\lambda(Q^2,s_{ij},m^2_k)}}\left(Q^2-m^2_k+s_{ij}\right)+Q^2+m^2_k-J^2_j\right]\frac{p_k}{2Q^2}\\
\nonumber & &+\left[\sqrt{\frac{\lambda(Q^2,J^2_j,m^2_k)}{\lambda(Q^2,s_{ij},m^2_k)}}\left(-Q^2-m^2_k+s_{ij}\right)+Q^2+m^2_k-J^2_j\right]\frac{p_i+p_j}{2Q^2}\\
\nonumber&=&p_k+(A_k-1)\; p_k+\frac{A_{ij}}{2}\;(p_i+p_j),\\
 \nonumber J_{j}&=&Q-J_{k},   \\ 
\nonumber&=&p_i+p_j-(A_k-1)\;p_k-\frac{A_{ij}}{2}\;(p_i+p_j),\\
  J_m&=&p_m, \quad (m\neq j,k)
\end{eqnarray}
which fulfil momentum conservation ($\sum\limits_{i=1}^n J_i=P$)
and the on-shell conditions ($J^2_j=m^2_{ij},\, J^2_l=m^2_l$ for $l\neq j$).

The inversion of the clustering is the same as in 
\Ref{Martini:2015fsa} with $y$ and $z$ as defined above.

In a $2\to 1$ clustering the clustering of two partons would be achieved by summing the two momenta while leaving all other momenta untouched. The difference between the $3\to 2$ clustering and the $2\to 1$ clustering as defined in \Eq{eq:fnorm} can be expressed as
\begin{eqnarray}\label{eq:ffnorm}
||J_j-(p_i+p_j)||
\nonumber &=&\max\left(\left|J^0_j-(p^0_i+p^0_j)\right|,\left|\vec{J}_j-(\vec{p}_i+\vec{p}_j)\right|\right)\\
&=&\max\left(\left|(A_k-1)\; p^0_k+\frac{A_{ij}}{2}\;(p^0_i+p^0_j)\right|,\left|(A_k-1)\; \vec{p}_k+\frac{A_{ij}}{2}\;(\vec{p}_i+\vec{p}_j)\right|\right).
\end{eqnarray}
For a given unresolved final state pair $i,j$ the spectator $k$ might be chosen such that it minimises $||J_j-(p_i+p_j)||$ given above.

\subsection{Final-state clustering with  initial-state spectator}
\label{sec:fim}
The phase space of $n+1$ massive partons can be
expressed as a phase space of $n$ particles convoluted with the dipole
phase space $dR_{ij,a}$:
\begin{equation}
  dR_{n+1}\left(p_a+p_b, p_1,\ldots,p_{n-1},p_j,p_i\right)
  =dR_{n}\left(xp_a+p_b, J_1,\ldots,J_{n-1},J_j\right)dR_{ij,a}.
\end{equation}
Using the \PSS it is again useful to use a
slightly different parameterisation compared to what has been used in 
\Ref{Martini:2015fsa}. We use the invariants
\begin{equation}
  Q^2=2xJ_j\cdot p_a, \quad
  s'_{ij}=2p_i\cdot p_j,\quad s_{ij}=s'_{ij}+m^2_i+m^2_j, \quad s'_{ia}=2p_i\cdot p_a 
\end{equation}
with
\begin{equation}
 x=\frac{Q^2}{Q^2-J^2_j+s_{ij}}, 
\end{equation}
and
\begin{equation}
 z=\frac{s'_{ia}}{Q^2-J^2_j+s_{ij}}.
\end{equation}
The phase space measure (after convolution with the PDFs and the flux factor) is given by
\begin{eqnarray}
\nonumber& &dx_Adx_B\frac{f_{a}\left(x_A\right)f_{b}\left(x_B\right)}{2x_Ax_Bs}dR_{n+1}\left(p_a+p_b, p_1,\ldots,p_{n-1},p_j,p_i\right)\\
\nonumber&=&dx_Adx_B\frac{f_{a}\left(\frac{x_A}{x}\right)f_{b}\left(x_B\right)}{2x_Ax_Bs}dR_{n}\left(xp_a+p_b,J_1,\ldots,J_{n-1},J_j\right)\frac{Q^2}{32\pi^3(Q^2-J^2_j+s_{ij})^2}\\
\nonumber&\times&\mkern-12mu d\phi\;ds'_{ij}\;ds'_{ia}\Theta\left(\phi\left(2\pi-\phi\right)\right)\Theta\left(\left(s'_{ia}-{s'_{ia}}^{-}\right)\left({s'_{ia}}^{+} -s'_{ia}\right)\right)\Theta\left(\left(s'_{ij}-{s'_{ij}}^{-}\right)\left({s'_{ij}}^{+} -s'_{ij}\right)\right).\\
\end{eqnarray}
The integration boundaries read
\begin{eqnarray*}
& z^{\mp}=\frac{s_{ij}+m^2_i-m^2_j\mp\sqrt{(s_{ij}-m^2_i-m^2_j)^2-4m^2_im^2_j}}{2s_{ij}},\quad {s'_{ia}}^{\mp}=z^{\mp}(Q^2-J^2_j+s_{ij}),&\\
&{s'_{ij}}^{-}=2m_im_j,\quad
{s'_{ij}}^{+}=\frac{1-x_A}{x_A} Q^2 +J^2_j -m^2_i -m^2_j.&
\end{eqnarray*}

The phase space parameterisation corresponds to the
clustering of $n+1$ partons to $n$ jets
\begin{eqnarray}\label{eq:ficlus}
  \label{eq:masslessMappingija1}
  J_j&=&\;p_i+p_j-\left(1-x\right)p_a,\\
  \label{eq:masslessMappingija2}
  J_m&=&\;p_m\quad (m\neq i,j)
\end{eqnarray}
which fulfils momentum conservation ($\sum\limits_{i=1}^n J_i= x p_a+ p_b$)
and the on-shell conditions ($J^2_j=m^2_{ij}$ and $J^2_l=m^2_l$ for $l\neq j$.).

The inversion of the clustering is the same as in 
\Ref{Martini:2015fsa} with $x$ and $z$ as defined above.

From \Eq{eq:ficlus} the difference between the $3\to 2$ clustering and the $2\to 1$ clustering defined in \Eq{eq:fnorm} follows as
\begin{equation}\label{eq:finorm}
||{J}_j - (p_i+p_j)||=(1-x)p^0_a.
\end{equation}
For a given unresolved final state pair $i,j$ the spectator $a$ might be chosen such that it minimises $||{J}_j - (p_i+p_j)||$ given above.

If both final state clusterings are possible the spectator should be chosen either from the final or the initial state according to the minimum of $||{J}_j - (p_i+p_j)||$ given in \Eq{eq:ffnorm} and \Eq{eq:finorm}.

\subsection{Initial-state clustering with final-state spectator }
\label{sec:ifm}
The phase space of $n$ massive and one massless parton can be
expressed as a convolution of the phase space of $n$ massive
jets and the dipole phase space $dR_{ia,k}$ for the emission
of an additional massless parton from the initial state with a massive
final-state spectator.  All statements from section \ref{sec:fim} can
be carried over by the replacements $a\rightarrow k$ and $j\rightarrow
a$, $m_i\rightarrow 0$ and $m_{ij}\rightarrow m_k$ (see also
\Ref{Martini:2015fsa}).

The phase space parameterisation corresponds to the
clustering of $n+1$ partons to $n$ jets
\begin{eqnarray}\label{eq:ifclus}
  \label{eq:masslessMappingiak1}
\nonumber  J_k&=&\;p_i+p_k-\left(1-x\right)p_a,\\
  \label{eq:masslessMappingiak2}
  J_m&=&\;p_m\quad (m\neq i,j)
\end{eqnarray}
which fulfils momentum conservation ($\sum\limits_{i=1}^n J_i= x p_a+ p_b$)
and the on-shell conditions ($J^2_l=m^2_l$, $l=1,\ldots,n$).

In a $2\to 1$ clustering the unresolved additional radiation
associated with the beam would be removed from the list of momenta
without changing the other final-state momenta.  So the difference
between the $3\to 2$ clustering defined in \Eq{eq:inorm} and the $2\to 1$
clustering can be expressed as (see \Eq{eq:ifclus})
\begin{eqnarray}\label{eq:ifnorm}
||{J}_k - p_k||=\max\left(\left|(1-x)p^0_a-p^0_i\right|,\left|(1-x)\vec{p}_a-\vec{p}_i\right|\right).
\end{eqnarray}
For a given unresolved final state parton $i$ the beam particle $a$
and the spectator $k$ might be chosen such that $||{J}_k - p_k||$ given above is
minimised.

\subsection{Initial-state clustering with  initial-state spectator}\label{sec:ii}
In case of initial-state clustering with an  initial-state spectator
the phase space can again be written as a convolution :
\begin{equation}\label{eq:psii}
  dR_{n+1}\left(p_a+p_b,p_1,\ldots,p_n,p_i\right)
  =dR_{n}\left(xp_a+p_b, J_1,\ldots,J_{n}\right)dR_{ia,b}.
\end{equation}

When employing a \PSS it is useful to express the integration after
convolution with the PDFs and the flux factor via the invariants
\begin{equation}
Q^2=2xp_a\cdot p_b,\quad 
s_{ia}=2p_i\cdot p_a, \quad
s_{ib}=2p_i\cdot p_b  
\end{equation}
 with 
 \begin{equation}
   x=\frac{Q^2}{Q^2+s_{ia}+s_{ib}}  
 \end{equation}
and
\begin{equation}
 v=\frac{s_{ia}}{Q^2+s_{ia}+s_{ib}}. 
\end{equation}
The phase space measure (after convolution with the PDFs and the flux factor) is given by
\begin{eqnarray}
\nonumber& &dx_Adx_B\frac{f_{a}\left(x_A\right)f_{b}\left(x_B\right)}{2x_Ax_Bs}dR_{n+1}\left(p_a+p_b, p_1,\ldots,p_{n-1},p_j,p_i\right)\\
\nonumber&=&dx_Adx_B\frac{f_{a}\left(\frac{x_A}{x}\right)f_{b}\left(x_B\right)}{2x_Ax_Bs}dR_{n}\left(xp_a+p_b,J_1,\ldots,J_{n}\right)\frac{Q^2}{32\pi^3(Q^2+s_{ia}+s_{ib})^2}\\
&&\times\; d\phi\;ds_{ia}\;ds_{ib}\Theta\left(\phi\left(2\pi-\phi\right)\right)\Theta\left(s_{ia}\left(s_{ia}^+-s_{ia}\right)\right)\Theta\left(s_{ik}\left(s_{ik}^+-s_{ik}\right)\right).
\end{eqnarray}
The integration boundaries read
\begin{eqnarray*}
&s_{ia}^+=\frac{1-x_A}{x_A} Q^2, \quad
s_{ib}^+=s_{ia}^+-s_{ia}.&
\end{eqnarray*}

The phase space parameterisation corresponds to the following
clustering of $n+1$ (massless/massive) partons to $n$ 
(massless/massive) jets:
\begin{equation}\label{eq:iiclus}
J_m=\;{\Lambda_{ia,b}}\;p_m, \quad m\neq i
\end{equation}
with
\begin{equation}
K=\;p_a+p_b-p_i,\quad \widetilde{K}=\;xp_{a}+p_b,
\end{equation}
and the Lorentz boost transforming $K$ into $\widetilde K$ given by
\begin{eqnarray}\label{eq:ltrsfii}
  {\left[\Lambda_{ia,b}\right]^{\mu}}_{\nu}&=& {g^{\mu}}_{\nu}
  -\frac{2\left(K+\widetilde{K}\right)^{\mu}\left(K+\widetilde{K}\right)_{\nu}}
  {\left(K+\widetilde{K}\right)^2}
  +\frac{2\widetilde{K}^{\mu}K_{\nu}}{K^2}.
\end{eqnarray}
The inverse boost is obtained by exchanging $K$ and $\widetilde K$.
All outgoing momenta $p_i$ are transformed to balance the transverse
momentum. Momentum conservation ($\sum\limits_{i=1}^n J_i=x p_a+p_b$)
and on-shell conditions ($J^2_l=m^2_l$, $l=1,\ldots,n$) are not
affected by the boost.

The inversion of the clustering is the same as in \Ref{Martini:2015fsa} with $x$ and $v$ as defined above.

The sum of the jet momenta in this $3\to 2$ clustering is given by
\begin{equation}
 \sum\limits_m {J}_m = \widetilde{K}
\end{equation}
while the sum of the partonic momenta is given by
\begin{equation}
 \sum\limits_{m\neq i} p_m = K.
\end{equation}
From \Eq{eq:iiclus} the influence of
the $3\to 2$ clustering with respect to the
$2\to 1$ clustering defined in \Eq{eq:inorm} follows as
\begin{equation}\label{eq:iinorm}
  ||\sum\limits_m {J}_m - \sum\limits_{m\neq i} p_m||=\max\left(\left|(1-x)p^0_a-p^0_i\right|,\left|(1-x)\vec{p}_a-\vec{p}_i\right|\right).
\end{equation}
For a given unresolved final state parton $i$ the beam particle $a$
and the spectator $b$ might be chosen such that  $||\sum J_m - \sum p_m||$ given above is
minimised.

If both initial-state clusterings are possible the beam particle and
the spectator (either from the final or the initial state) should be
chosen according to the minimum of the quantities given by \Eq{eq:ifnorm} and \Eq{eq:iinorm}.

\section{Additional figures}

\subsection{Comparison of $3\to 2$ and $2\to 1$ clustering}
\label{sec:comparison-clustering}
\begin{figure}[htbp]
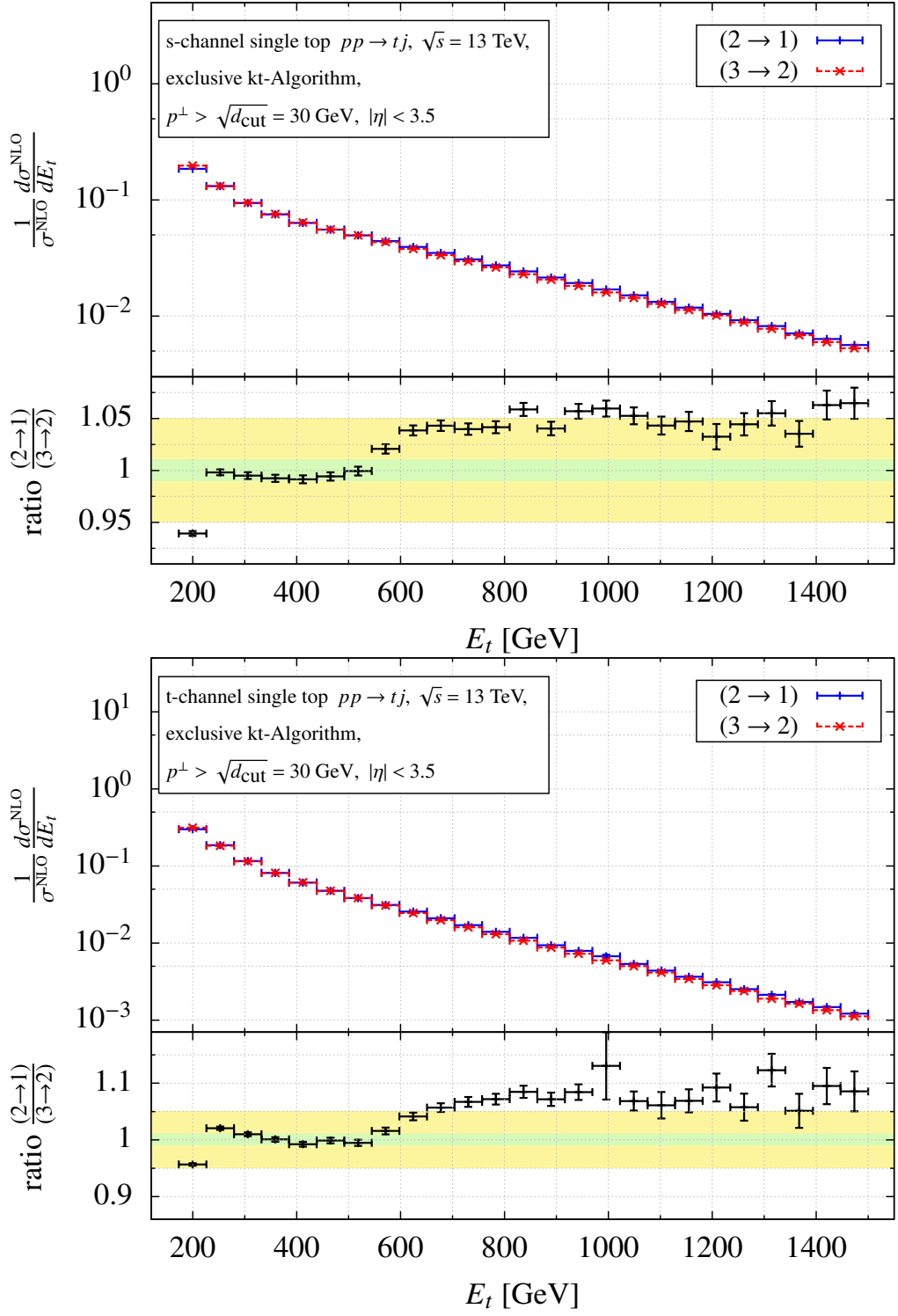

  \begin{flushright}
    \leavevmode
    \includegraphics[height=0.47\textheight]{{{%
          sgtsKT3-2ycut30compjetE1-200x1e7-crop}}}
    \includegraphics[height=0.47\textheight]{{{%
          sgttKT3-2ycut30compjetE1-200x1e7-crop}}}
    \caption{Energy distributions of the top-tagged jet 
      from $s$- and $t$-channel single top-quark production at NLO with
      $3\to 2$ (red, dashed) and $2\to 1$ (blue, solid) jet clusterings.}
    \label{fig:jetcom_sgten1app}
  \end{flushright}
\end{figure}
\begin{figure}[htbp]
  \begin{flushright}
    \leavevmode
    \includegraphics[height=0.47\textheight]{{{%
          sgtsKT3-2ycut30compjetET2-200x1e7-crop}}}
    \includegraphics[height=0.47\textheight]{{{%
          sgttKT3-2ycut30compjetET2-200x1e7-crop}}}
    \caption{Transverse energy distributions of the light jet 
      from $s$- and $t$-channel single top-quark production at NLO with
      $3\to 2$ (red, dashed) and $2\to 1$ (blue, solid) jet clusterings.}
    \label{fig:jetcom_sgtentr2app}
  \end{flushright}
\end{figure}
\begin{figure}[htbp]
  \begin{flushright}
    \leavevmode
    \includegraphics[height=0.47\textheight]{{{%
          sgtsKT3-2ycut30compjetET1-200x1e7-crop}}}
    \includegraphics[height=0.47\textheight]{{{%
          sgttKT3-2ycut30compjetE2-200x1e7-crop}}}
    \caption{(Transverse) Energy distributions of the top-tagged and light jet 
      from $s$- and $t$-channel single top-quark production at NLO with
      $3\to 2$ (red, dashed) and $2\to 1$ (blue, solid) jet clusterings.}
    \label{fig:jetcom_sgtenapp}
  \end{flushright}
\end{figure}

\begin{figure}[htbp]
  \begin{flushright}
    \includegraphics[height=0.47\textheight]{{{%
          sgtsKT3-2ycut30compjetETA2-200x1e7-crop}}}
    \includegraphics[height=0.47\textheight]{{{%
          sgttKT3-2ycut30compjetETA1-200x1e7-crop}}}
    \caption{Pseudo rapidity distributions for the top-tagged and light jet 
      from $s$- and $t$-channel single top-quark production at NLO with
      $3\to 2$ (red, dashed) and $2\to 1$ (blue, solid) jet clusterings.}
    \label{fig:jetcom_sgtanapp}
  \end{flushright}
\end{figure}
\begin{figure}[htbp]
  \begin{flushright}
    \includegraphics[height=0.47\textheight]{{{%
          sgtsKT3-2ycut30compjetCTH1-200x1e7-crop}}}
    \includegraphics[height=0.47\textheight]{{{%
          sgttKT3-2ycut30compjetCTH1-200x1e7-crop}}}
    \caption{Polar angle distributions for the top-tagged jet 
      from $s$- and $t$-channel single top-quark production at NLO with
      $3\to 2$ (red, dashed) and $2\to 1$ (blue, solid) jet clusterings.}
    \label{fig:jetcom_sgtpr1app}
  \end{flushright}
\end{figure}
\begin{figure}[htbp]
  \begin{flushright}
    \includegraphics[height=0.47\textheight]{{{%
          sgtsKT3-2ycut30compjetCTH2-200x1e7-crop}}}
    \includegraphics[height=0.47\textheight]{{{%
          sgttKT3-2ycut30compjetCTH2-200x1e7-crop}}}
    \caption{Polar angle distributions for the light jet 
      from $s$- and $t$-channel single top-quark production at NLO with
      $3\to 2$ (red, dashed) and $2\to 1$ (blue, solid) jet clusterings.}
    \label{fig:jetcom_sgtpr2app}
  \end{flushright}
\end{figure}

\subsection{NLO corrections to differential distributions and
  k-factors}
\label{sec:kfactors}

\begin{figure}[htbp]
  \begin{center}
    \leavevmode
    \includegraphics[height=0.325\textheight]{{{%
          sgtsKT3-2ycut30compmumnspecE2-200x1e7nonorm-crop}}}
    \includegraphics[height=0.325\textheight]{{{%
          sgtsKT3-2ycut30compmumnspecETA2-200x1e7nonorm-crop}}}    
    \includegraphics[height=0.325\textheight]{{{%
          sgtsKT3-2ycut30compmumnspecETA1-200x1e7nonorm-crop}}}
    \caption{Unnormalised distributions and k-factors for $s$-channel
      single top-quark production using the exclusive event
      definition.}  
    \label{fig:kfac_sgtsapp}
  \end{center}
\end{figure}

\begin{figure}[htbp]
  \begin{center}
    \leavevmode
    \includegraphics[height=0.325\textheight]{{{%
          sgttKT3-2ycut30compmumnspecE2-200x1e7nonorm-crop}}}
    \includegraphics[height=0.325\textheight]{{{%
          sgttKT3-2ycut30compmumnspecETA2-200x1e7nonorm-crop}}}    
    \includegraphics[height=0.325\textheight]{{{%
          sgttKT3-2ycut30compmumnspecETA1-200x1e7nonorm-crop}}}
    \caption{Same as \Fig{fig:kfac_sgtsapp} but for the $t$-channel.}  
    \label{fig:kfac_sgttapp}
  \end{center}
\end{figure}

\begin{figure}[htbp]
  \begin{center}
    \leavevmode
    \includegraphics[height=0.325\textheight]{{{%
          sgtsKT3-2ycut30compmumnspecincE2-200x1e7nonorm-crop}}}
    \includegraphics[height=0.325\textheight]{{{%
          sgtsKT3-2ycut30compmumnspecincETA2-200x1e7nonorm-crop}}}    
    \includegraphics[height=0.325\textheight]{{{%
          sgtsKT3-2ycut30compmumnspecincETA1-200x1e7nonorm-crop}}}
    \caption{Same as \Fig{fig:kfac_sgtsapp}  but for the inclusive event 
      definition. }
    \label{fig:kfac_sgtsincapp}
  \end{center}
\end{figure}  

\begin{figure}[htbp]
  \begin{center}
    \leavevmode
    \includegraphics[height=0.325\textheight]{{{%
          sgttKT3-2ycut30compmumnspecincE2-200x1e7nonorm-crop}}}
    \includegraphics[height=0.325\textheight]{{{%
          sgttKT3-2ycut30compmumnspecincETA2-200x1e7nonorm-crop}}}    
    \includegraphics[height=0.325\textheight]{{{%
          sgttKT3-2ycut30compmumnspecincETA1-200x1e7nonorm-crop}}}
    \caption{Same as \Fig{fig:kfac_sgttapp}  but for the inclusive event 
      definition. }
    \label{fig:kfac_sgttincapp}
  \end{center}
\end{figure}

\clearpage
\providecommand{\href}[2]{#2}\begingroup\raggedright\endgroup

\end{document}